\documentclass[a4paper,11pt]{article}

\usepackage{jcappub} 

\usepackage[T1]{fontenc} 

\usepackage{amsmath,amssymb,amsfonts}
\usepackage{mathtools}
\usepackage{bm}
\usepackage{bbm}

\usepackage{graphicx}
\usepackage{xcolor}
\usepackage{caption}
\usepackage{placeins}

\usepackage{comment}

\usepackage{natbib}
\usepackage{aas_macros}

\usepackage{hyperref}
\hypersetup{unicode=true}

\usepackage{amsthm}
\usepackage{tikz}
\usetikzlibrary{arrows.meta,positioning,calc,fit,cd,decorations.pathmorphing,intersections}

\theoremstyle{definition}

\theoremstyle{plain}

\theoremstyle{plain}

\theoremstyle{plain}

\theoremstyle{remark}

\theoremstyle{remark}

\theoremstyle{remark}

\def\pd{{\rm d}}

\def\deltadir{\delta_{\rm D}}

\title{Joint State-and-Dynamics Inference for Galaxy Population Evolution on an Effective Manifold}

\author[a,b,1]{Tsutomu T. Takeuchi,\note{Corresponding author.}}
\author[a,c]{Ryusei R. Kano}

\affiliation[a]{Division of Particle and Astrophysical Science, Nagoya University, Furo-cho, Chikusa-ku, Nagoya, Aichi 464--8602, Japan}
\affiliation[b]{The Research Center for Statistical Machine
Learning, the Institute of Statistical Mathematics, 10--3 Midori-cho, Tachikawa, Tokyo 190--8562, Japan}
\affiliation[c]{The Institute for Astronomy (IfA), School of Physics and Astronomy, the University of Edinburgh, Royal Observatory, Edinburgh, EH9 3HJ, UK}

\emailAdd{tsutomu.takeuchi.ttt@gmail.com}

\abstract{
Galaxy surveys provide noisy, incomplete, selection-affected population
snapshots rather than complete evolutionary histories. We formulate galaxy evolution as joint inference of effective physical states, their population intensity, and cross-epoch dynamics. 
An effective state is a coarse-grained description defined by predictive adequacy and distinguished from measured variables and learned representations. 
The intrinsic population is a finite non-negative measure evolving through one-galaxy transport, source and removal terms, and nonlocal two-to-one merger jumps.
A sub-probability observational kernel maps this population to survey catalogs and supplies the likelihood for state inference.
Projected luminosity and stellar-mass functions generally do not obey closed dynamics and cannot identify the underlying channels alone. 
In a restricted IllustrisTNG proof of concept, a finite-time non-merger kernel and reduced merger operator are calibrated on training root merger trees and evaluated on held-out lineages. 
Dynamical propagation improves the normalized later mass--sSFR distribution relative to a static baseline, while the merger channel captures number loss from disappearing
secondary progenitors. 
Under noisy and censored mock observations, the propagated prior improves posterior-stacked reconstruction of the latent population relative to a frozen-static prior. 
The framework provides an
observation-directed, simulation-assisted route to galaxy
state-and-dynamics inference.
}

\begin{document}

\maketitle

\flushbottom

\section{Introduction}
\label{sec:introduction}

Galaxy formation and evolution arise from the coupled action of hierarchical assembly, gas accretion and cooling, star formation, feedback, chemical enrichment, environmental processing, and galaxy mergers. These processes are not observed as complete trajectories of
individual systems. 
Surveys instead provide noisy, incomplete, and selection-affected population snapshots at a finite set of cosmic epochs. Moreover, many standard galaxy properties, including stellar mass, star-formation rate, metallicity, dust attenuation, and gas content, are themselves inferred from photometric or spectroscopic data through physical and statistical models \citep{2013ARA&A..51..393C}. 
The central observational problem is therefore inverse: one must infer the evolving galaxy states and the population-level transitions that could have generated the measured catalogs.

The objective of this work is to formulate a joint state-and-dynamics inference problem that embeds projected statistics such as luminosity functions, stellar-mass functions, and color distributions within a common cross-epoch generative model.
Let
\begin{align}
    \mathcal{D}_{\rm obs}
    &=
    \left\{
        \bigl(\bm{x}_a,t_a\bigr)
    \right\}_{a=1}^{N_{\rm gal}}
    \notag
\end{align}
denote an observational catalog, where $\bm{x}_a$ is the measured data
vector for galaxy $a$ and $t_a$ is its cosmic epoch. The scientific
target is schematically
\begin{align}
    p\!\left(
        \{\theta_a\}_{a=1}^{N_{\rm gal}},
        \rho,
        \bm{\beta}
        \mid
        \mathcal{D}_{\rm obs}
    \right),
    \label{eq:intro_inference_target}
\end{align}
where $\theta_a$ is an effective physical state, $\rho(\theta,t)$ is the population intensity density on the corresponding state space, and $\bm{\beta}$ denotes parameters of the dynamical and observation models.
Equation~\eqref{eq:intro_inference_target} defines the inferential objective relative to an explicit state model and observation process, with uncertainty and non-identifiability retained as part of the result.

The word \emph{state} is used operationally. We distinguish the measured variables $\bm{x}$, an algorithm-dependent representation $\bm{z}=\Psi(\bm{x})$, and the effective physical state $\theta$. 
In general, $\bm{z}\neq\theta$. 
A learned representation may support visualization, compression, or inference, whereas the modeled physical variables are specified through the effective-state definition and its predictive task.
We use the term \emph{galaxy manifold} for the effective state space $\mathcal{M}$, continuing a long tradition of describing the structured organization of galaxy populations in multivariate spaces \citep{1973A&A....23..259B,1992ASSL..178..337D,
takeuchi2025applications,2026Entrp..28..288T}. 
Here, $\mathcal{M}$ is neither a unique hidden geometry nor automatically the output of a manifold-learning algorithm. 
It is a coarse-grained, model-dependent state space chosen for a specified temporal resolution and prediction task. 
A proposed $\theta$ is adequate only insofar as omitted histories, environments, halo variables, or internal properties do not retain substantial predictive information about future transitions. 
Otherwise, the state must be enlarged.

We impose cross-epoch consistency by representing the galaxy population as a finite non-negative measure on $\mathcal{M}$, with density $\rho(\theta,t)$ when a density representation is appropriate, and evolving it according to
\begin{align}
    \frac{\partial \rho(\theta,t)}{\partial t}
    &=
    -\nabla_{\mathcal{M}}\!\cdot
    \left[
        \rho(\theta,t)\bm{v}(\theta,t)
    \right]
    +B(\theta,t)
    -\Lambda(\theta,t)\rho(\theta,t)
    +J_t[\rho](\theta).
    \label{eq:intro_master}
\end{align}
The transport field $\bm{v}$ describes one-galaxy evolution that can be
represented as continuous motion through the effective state space;
$B$ and $\Lambda$ describe entry into and removal from the intrinsic
population; and $J_t$ represents interaction-driven jumps. 
Galaxy mergers require a distinct jump structure because they couple two progenitor states and produce one remnant. 
The resulting operator is therefore nonlinear in the population intensity and changes galaxy number. 
Equation~\eqref{eq:intro_master} provides a reduced population-level model whose coefficients summarize the consequences of gravitational, hydrodynamic, and subgrid processes at the adopted coarse-graining.

The state dynamics are connected to observations by a time-dependent sub-probability kernel $\mathcal{K}_{\rm obs}(\pd\bm{x}\mid\theta,t)$, which may include
intrinsic scatter, measurement uncertainty, selection, censoring, and instrumental response. 
If this kernel has density $k_{\rm obs}(\bm{x}\mid\theta,t)$, the predicted observational intensity
is
\begin{align}
    \lambda_{\rm obs}(\bm{x},t)
    &=
    \int_{\mathcal{M}}
    k_{\rm obs}(\bm{x}\mid\theta,t)
    \rho(\theta,t)
    \,\pd V_{\mathcal{M}}.
    \label{eq:intro_forward_model}
\end{align}
This forward model makes clear why projected statistics generally do not
identify the underlying dynamics uniquely: distinct state distributions
and combinations of transport, removal, and merger processes may produce
similar observable marginals. Conversely, in state inference the
observational kernel supplies the likelihood, while the solution of the
transport--jump model supplies a time-dependent population prior. The
dynamical theory thus links different epochs and restricts posterior
state reconstructions to a common evolutionary model.

The structure of Equation~\eqref{eq:intro_master} draws on three
established lines of work. Hierarchical theories of halo assembly and
semi-analytic galaxy formation provide statistical and physical forward
models for galaxy histories
\citep{1975A&A....45..365G,1974ApJ...187..425P,
1991ApJ...379..440B,1991ApJ...379...52W}. Kinetic merger models represent
binary interactions through nonlocal gain and loss, with classical
Smoluchowski coagulation as the scalar additive limit
\citep{smoluchowski1917versuch,1978ApJ...220..390S,
1978ApJ...223L..59S}. Continuity-equation approaches describe smooth
stellar-mass growth and population transfer as advective fluxes
\citep{2008ApJ...680...41D,2010ApJ...721..193P}. Related analyses of downsizing and merger histories show that observed evolutionary trends can emerge from mass-dependent efficiencies, quenching, and multi-progenitor assembly rather than from a single projected coordinate
\citep{1996AJ....112..839C,2006MNRAS.372..933N,
2008MNRAS.388.1792N}. 
The present framework combines these structural elements on a multivariate effective state space and separates the rate of a merger from the conditional distribution of its remnant state.

Numerical simulations provide calibration and validation data for specified effective-state and dynamical models.
Cosmological hydrodynamic simulations provide controlled trajectories, descendant links, and merger histories with which one can test state adequacy, estimate finite-time operators, and validate inference when the
underlying histories are known \citep{2018MNRAS.473.4077P,2019ComAC...6....2N}. 
When the likelihood of a specified state-and-dynamics model is intractable, SBI provides a computational route to its parameter posterior \citep{2019MNRAS.488.4440A,2020PNAS..11730055C}. 

We develop the framework at complementary analytic and numerical levels.
First, luminosity functions and related projected observables are used to show explicitly that projection generally destroys dynamical closure and identifiability. 
The Schechter family is studied as a candidate effective invariant family whose tangency conditions constrain projected evolution \citep{1976ApJ...203..297S}. 
Downsizing is formulated through fluxes into quenched regions and state-dependent residence times, incorporating the multiple channels that govern the evolving star-forming population.
Second, a controlled IllustrisTNG proof of concept uses the interpretable two-dimensional mass--sSFR state $\theta=(m,s)$. 
A finite-time non-merger kernel and a reduced operator with explicit merger channels are calibrated on training root merger trees and applied to a held-out population. 
The two finite-time evolution models improve prediction of the later distribution over a static baseline, the explicit merger channel recovers the number-changing balance associated with disappearing secondary progenitors, and the dynamically propagated prior improves reconstruction of the latent later population from noisy and censored mock observations. 

The main contributions can therefore be summarized compactly. We define
an effective galaxy state through predictive adequacy and distinguish it
from both measured variables and learned representations. 
We formulate population evolution as transport, source and removal terms, and binary-merger jumps on a finite non-negative measure, including estimable finite-time operators. 
We connect the resulting dynamical model to selection-affected catalogs and Bayesian state-and-dynamics inference, using simulations for calibration and controlled validation and, when needed, simulation-based methods for posterior computation.
Finally, we provide analytic projection diagnostics and a held-out numerical test showing what the formalism enables in practice.

The remainder of this paper is organized as follows.
Section~\ref{sec:effective_states} defines effective states, observational representations, and population measures.
Section~\ref{sec:population_dynamics} develops one-galaxy transport and finite-time transition kernels, while Section~\ref{sec:merger_chapter} formulates binary mergers as nonlocal jump processes.
Section~\ref{sec:state_inference} constructs the observational forward model and the joint inference problem.
Section~\ref{sec:applications} examines projected observables, the Schechter family, downsizing, and identifiability.
Section~\ref{sec:calibration} discusses constrained parameterization, regularization, calibration, and computational inference.
Section~\ref{sec:proof_of_concept} presents the IllustrisTNG proof of concept, and Section~\ref{sec:conclusion} summarizes the results and
outlook.

\section{Effective Galaxy States, Observational Representations,
and Population Measures}
\label{sec:effective_states}

The term \emph{galaxy manifold} has been used for the structured
organization of galaxy populations in multivariate spaces
\citep{1973A&A....23..259B,1992ASSL..178..337D,
takeuchi2025applications,2026Entrp..28..288T}. Here it has a restricted,
operational meaning. 
The manifold $\mathcal{M}$ is a coarse-grained, model-dependent state space chosen for a specified temporal resolution and prediction task.

\subsection{Effective galaxy states and predictive adequacy}
\label{subsec:effective_state}

Let $\xi_t\in\mathcal{S}$ denote a detailed physical state at cosmic time
$t$. In a simulation, $\xi_t$ may include resolved internal variables,
halo and environmental properties, and assembly history, most of which
are inaccessible observationally. An effective state is a specified
coarse-graining
\begin{align}
    \theta_t
    &=
    C(\xi_t),
    \qquad
    C:\mathcal{S}\rightarrow\mathcal{M},
    \label{eq:effective_state_coarse_graining}
\end{align}
chosen for a particular temporal resolution and prediction task. The map
$C$ is part of the model: it may be physically prescribed, calibrated
with simulations, or informed by observations, but it is not assumed to
be unique or fundamental.

A proposed state is adequate when omitted information does not retain
substantial predictive power for the transition being modeled. If
$\mathcal{I}^{\rm extra}_t$ denotes resolved information excluded from
$\theta_t$, this requirement may be written
\begin{align}
    p\!\left(
        \theta_{t+\Delta t}
        \mid
        \theta_t,
        \mathcal{I}^{\rm extra}_t,
        t
    \right)
    &\simeq
    p\!\left(
        \theta_{t+\Delta t}
        \mid
        \theta_t,
        t
    \right).
    \label{eq:predictive_adequacy}
\end{align}
This is an approximate closure condition, not a claim of microscopic
Markovianity. It can be tested on held-out trajectories by asking whether
transition residuals still depend on omitted variables such as halo mass,
environment, central or satellite status, or earlier history. A
persistent dependence indicates that the state should be enlarged,
whereas variables that do not improve prediction need not be retained.
The proof of concept in Section~\ref{sec:proof_of_concept} adopts the interpretable two-dimensional mass--sSFR state $\theta=(m,s)$, whose predictive adequacy can be assessed using the validation criteria developed here.

For the geometric formulation, $\mathcal{M}$ is taken to be an $m$-dimensional smooth manifold with Riemannian metric $g$ and volume element
\begin{align}
    \pd V_{\mathcal{M}}
    &=
    \sqrt{\det g(\theta)}
    \,\pd\theta^1\cdots\pd\theta^m.
    \label{eq:manifold_volume}
\end{align}
The metric and its scale are themselves components of the effective
model. Discrete attributes may be represented by separate strata or by
conditioning the dynamics on an additional discrete label.

\subsection{Detailed states, observations, and learned representations}
\label{subsec:state_observation_representation}

Four spaces must be distinguished:
\begin{align}
    \xi\in\mathcal{S}
    &\xrightarrow{\ C\ }
    \theta\in\mathcal{M}
    \xrightarrow{\ \mathcal{K}_{\rm obs}\ }
    \bm{x}\in X
    \xrightarrow{\ \Psi\ }
    \bm{z}\in Z.
    \notag
\end{align}
The first arrow is the coarse-graining in
Equation~\eqref{eq:effective_state_coarse_graining}. The second is
usually stochastic and describes how a physical state generates measured
data. The third is a deterministic or stochastic representation map
constructed from those data.

The observational vector $\bm{x}$ may contain fluxes, spectra, image
quantities, or derived estimates such as stellar mass and star-formation
rate. Derived estimates remain observational variables because they
inherit the assumptions and uncertainties of their measurement model.
The observation process is represented by
$\mathcal{K}_{\rm obs}(\pd\bm{x}\mid\theta,t)$, which may include
intrinsic scatter, measurement noise, censoring, selection, and
instrumental response.

A learned representation is written
\begin{align}
    \bm{z}
    &=
    \Psi(\bm{x}),
    \qquad
    \Psi:X\rightarrow Z,
    \label{eq:representation_map}
\end{align}
but in general $\bm{z}\neq\theta$, even when the two spaces have the
same dimension. The representation depends on the input variables,
preprocessing, learning objective, algorithm, and hyperparameters;
the effective state is defined by physical interpretation and predictive
adequacy. A representation can inform state construction, but connecting
the two requires an additional calibration model, such as
$q(\theta\mid\bm{z},t)$, rather than an identification supplied by the
embedding itself.

\subsection{Population measures and pushforward relations}
\label{subsec:population_measures}

At time $t$, let $\mathsf{N}_t$ be a finite non-negative measure on
$\mathcal{M}$. For a measurable set $A\subset\mathcal{M}$,
$\mathsf{N}_t(A)$ is the expected galaxy count, or corresponding number
density, in that state region. When the measure has a density with
respect to $\pd V_{\mathcal{M}}$,
\begin{align}
    \mathsf{N}_t(\pd\theta)
    &=
    \rho(\theta,t)\,\pd V_{\mathcal{M}},
    \label{eq:physical_population_measure}\\
    N_{\rm tot}(t)
    &=
    \mathsf{N}_t(\mathcal{M})
    =
    \int_{\mathcal{M}}
    \rho(\theta,t)\,\pd V_{\mathcal{M}}.
    \label{eq:total_population_intensity}
\end{align}
The intensity $N_{\rm tot}$ need not be conserved, notably because two
merger progenitors are replaced by one remnant. When a normalized state
distribution is needed, it is defined separately as
\begin{align}
    \pi_t(\pd\theta)
    &=
    \frac{\mathsf{N}_t(\pd\theta)}{N_{\rm tot}(t)},
    \qquad
    N_{\rm tot}(t)>0.
    \notag
\end{align}
Thus, $\rho$ is not assumed to integrate to unity.

The observation process acts on the physical population through the
sub-probability kernel
$\mathcal{K}_{\rm obs}(\pd\bm{x}\mid\theta,t)$. Its total mass
$\mathcal{K}_{\rm obs}(X\mid\theta,t)$ is the probability that a galaxy
in state $\theta$ enters the observed sample. The resulting
observational intensity measure is
\begin{align}
    \mathsf{M}_t(E)
    &=
    \int_{\mathcal{M}}
    \mathcal{K}_{\rm obs}(E\mid\theta,t)
    \mathsf{N}_t(\pd\theta),
    \label{eq:observational_intensity_measure}
\end{align}
for measurable $E\subset X$. If the kernel and population measure admit
densities, then
\begin{align}
    \lambda_{\rm obs}(\bm{x},t)
    &=
    \int_{\mathcal{M}}
    k_{\rm obs}(\bm{x}\mid\theta,t)
    \rho(\theta,t)
    \,\pd V_{\mathcal{M}},
    \label{eq:observational_intensity_density}
\end{align}
where $\lambda_{\rm obs}$ is the expected catalog intensity. The
normalized observational measure
\begin{align}
    \mu_t(E)
    &=
    \frac{\mathsf{M}_t(E)}{\mathsf{M}_t(X)},
    \qquad
    \mathsf{M}_t(X)>0.
    \label{eq:normalized_observational_measure}
\end{align}
contains the distributional shape but not the total observed count; number evolution must therefore be retained separately whenever it is scientifically relevant.

For a representation map $\Psi:X\rightarrow Z$, the corresponding representation measure is the pushforward
\begin{align}
    \nu_t
    &=
    \Psi_{\#}\mu_t.
    \label{eq:representation_pushforward}
\end{align}
This relation formalizes what representation learning does to the observed population, but it does not identify $\nu_t$ with the intrinsic population measure $\mathsf{N}_t$. 
Information not preserved by $\Psi$ is lost, and $\nu_t$ remains dependent on the observational variables and learning procedure. 
A physical interpretation therefore requires calibration back to $\theta$ through the observation model or an explicit conditional relation. 
The population dynamics in the following sections are accordingly formulated on $\mathsf{N}_t$, or on $\rho(\theta,t)$ when a density exists; observational and representation measures enter through the forward and inverse problems. 
A fuller measure-theoretic treatment is given in \cite{takeuchi2026measuretheoretic}.

\section{Population Dynamics on the Effective Galaxy Manifold}
\label{sec:population_dynamics}

The preceding section defined the effective state space $\mathcal{M}$
and the population measure $\mathsf{N}_t$. We now specify the evolution
of the one-galaxy channel, in which one effective state is transformed
into another without an explicit interaction between two population
members. Galaxy mergers are treated separately as binary jumps in
Section~\ref{sec:merger_chapter}.

\subsection{Continuous transport and population balance}
\label{subsec:continuous_transport}

Suppose that $\mathsf{N}_t$ has density $\rho(\theta,t)$ with respect to
$\pd V_{\mathcal{M}}$. A tangent field $\bm{v}(\theta,t)$ represents
local drift through the effective state space, while
$B(\theta,t)\geq0$ and $\Lambda(\theta,t)\geq0$ denote intrinsic source
intensity and state-dependent removal rate. The continuous population
balance is
\begin{align}
    \frac{\partial\rho(\theta,t)}{\partial t}
    &=
    -\nabla_{\mathcal{M}}\!\cdot
    \left[
        \rho(\theta,t)\bm{v}(\theta,t)
    \right]
    +B(\theta,t)
    -\Lambda(\theta,t)\rho(\theta,t).
    \label{eq:continuous_population_dynamics}
\end{align}
The divergence term redistributes existing galaxies, $B$ adds members,
and $\Lambda\rho$ removes them. Which process belongs to which channel
depends on the state and population definition. For example, quenching
is transport when both star-forming and quenched systems belong to
$\mathcal{M}$, but removal when the model is restricted to the
star-forming population. These terms describe the intrinsic population;
survey selection and incompleteness enter separately through
$\mathcal{K}_{\rm obs}$.

Under a no-flux boundary condition, integration of
Equation~\eqref{eq:continuous_population_dynamics} gives
\begin{align}
    \frac{\pd N_{\rm tot}(t)}{\pd t}
    &=
    \int_{\mathcal{M}}
    \left[
        B(\theta,t)
        -\Lambda(\theta,t)\rho(\theta,t)
    \right]
    \,\pd V_{\mathcal{M}}.
    \label{eq:continuous_number_balance}
\end{align}
Thus, internal one-galaxy transport conserves the number of retained
objects; count changes arise from sources, removal, and the merger
channel introduced below. If the state is reduced to stellar mass,
$\bm{v}=\dot M_\ast$, and $B=\Lambda=0$, the same equation becomes
\begin{align}
    \frac{\partial\rho(M_\ast,t)}{\partial t}
    +
    \frac{\partial}{\partial M_\ast}
    \left[
        \dot M_\ast(M_\ast,t)\rho(M_\ast,t)
    \right]
    &=0,
    \notag
\end{align}
recovering the usual continuity description of smooth stellar-mass
growth \citep{2008ApJ...680...41D,2010ApJ...721..193P}.

\subsection{Finite-time transition kernels}
\label{subsec:finite_time_transition}

Surveys and simulations usually provide populations at discrete epochs,
so the one-galaxy evolution is also described by a finite-time kernel
\begin{align}
    \mathcal{Q}^{(0)}_{t,s}
    \left(
        \pd\theta'\mid\theta
    \right),
    \qquad
    \mathcal{Q}^{(0)}_{t,s}
    \left(
        \mathcal{M}\mid\theta
    \right)
    &=1,
    \qquad
    t>s.
    \notag
\end{align}
The superscript $(0)$ denotes the non-merger channel, conditional here
on retention in the modeled population. The kernel need not collapse
to a deterministic trajectory: its width may represent stochastic
histories, omitted state variables, finite snapshot spacing, or
heterogeneity among galaxies assigned to the same coarse-grained state.
When source and removal terms are absent, it propagates the population measure
according to
\begin{align}
    \mathsf{N}_t(A)
    &=
    \int_{\mathcal{M}}
    \mathcal{Q}^{(0)}_{t,s}
    \left(
        A\mid\theta
    \right)
    \mathsf{N}_s(\pd\theta),
    \label{eq:finite_time_measure_evolution}
\end{align}
for measurable $A\subset\mathcal{M}$. Removal or boundary escape can be
represented by a sub-probability kernel, with the missing mass recording
loss from the population; source contributions can be added by
propagating the injected measure from its time of entry.

Predictive adequacy of the chosen effective state implies approximate
composition across intermediate epochs:
\begin{align}
    \mathcal{Q}^{(0)}_{t,s}
    \left(
        A\mid\theta
    \right)
    &\simeq
    \int_{\mathcal{M}}
    \mathcal{Q}^{(0)}_{t,u}
    \left(
        A\mid\vartheta
    \right)
    \mathcal{Q}^{(0)}_{u,s}
    \left(
        \pd\vartheta\mid\theta
    \right),
    \qquad
    s<u<t.
    \label{eq:chapman_kolmogorov_effective}
\end{align}
Systematic failure of
Equation~\eqref{eq:chapman_kolmogorov_effective} indicates that the
state description, the temporal resolution, or both are inadequate.

For a partition $\mathcal{M}=\bigcup_i C_i$, define the cell transition
probability and cell count by
\begin{align}
    Q^{(0)}_{ji}(t,\Delta t)
    &=
    \Pr\!\left(
        \theta_{t+\Delta t}\in C_j
        \mid
        \theta_t\in C_i,
        \text{no resolved merger}
    \right),
    \label{eq:discrete_nonmerger_transition}\\
    n_j(t+\Delta t)
    &=
    \sum_{i=1}^{N_{\rm cell}}
    Q^{(0)}_{ji}(t,\Delta t)n_i(t),
    \qquad
    n_i(t)=\mathsf{N}_t(C_i).
    \label{eq:discrete_nonmerger_update}
\end{align}
This matrix is a directly estimable finite-time coarse-grained model at the temporal resolution of the matched snapshot data and is used for the non-merger channel in the controlled numerical demonstration.

\subsection{Transport--jump decomposition}
\label{subsec:transport_jump_decomposition}

Collecting the one-galaxy contributions into
\begin{align}
    \mathcal{T}_t[\rho](\theta)
    &=
    -\nabla_{\mathcal{M}}\!\cdot
    \left[
        \rho(\theta,t)\bm{v}(\theta,t)
    \right]
    +B(\theta,t)
    -\Lambda(\theta,t)\rho(\theta,t),
    \notag
\end{align}
the full population equation is
\begin{align}
    \frac{\partial\rho(\theta,t)}{\partial t}
    &=
    \mathcal{T}_t[\rho](\theta)
    +J_t[\rho](\theta).
    \label{eq:transport_jump_decomposition}
\end{align}
The distinction is structural rather than geometric. A one-galaxy
kernel may have broad support and move a system across a large distance
in state space, but it remains a one-to-one transition retaining one
galaxy identity. A merger instead couples two progenitors and produces
one remnant; its rate depends on population pairs, its gain-and-loss
operator is generally quadratic in $\rho$, and it changes galaxy number.
Section~\ref{sec:merger_chapter} constructs $J_t[\rho]$ by separating the
merger rate from the conditional remnant-state distribution.

\section{Galaxy Mergers as Nonlocal Jump Processes}
\label{sec:merger_chapter}

Section~\ref{sec:population_dynamics} described the one-galaxy channel,
in which one system is retained while its effective state changes. A
binary merger instead has the population-level structure
\begin{align}
    (\theta_1,\theta_2)
    &\longrightarrow
    \theta.
    \notag
\end{align}
It removes two progenitors, creates one remnant, and depends on pairs
drawn from the population. Its contribution is therefore generally
quadratic in the population intensity. Classical coagulation theory
provides the corresponding gain-and-loss structure
\citep{smoluchowski1917versuch,1978ApJ...220..390S,
1978ApJ...223L..59S}; here it is extended from a single additive
variable to a multivariate effective state.

\subsection{Binary merger operator}
\label{subsec:binary_merger_operator}

Let $R(\theta_1,\theta_2;t)\geq0$ be the effective merger-rate kernel.
For a galaxy in state $\theta_1$, 
\begin{align}
    R(\theta_1,\theta_2;t) \rho(\theta_2,t)\pd V_{\mathcal M,2}
    \notag
\end{align}
is the instantaneous rate of merging with a partner near $\theta_2$. 
We assume exchange symmetry,
\begin{align}
    R(\theta_1,\theta_2;t)
    &=
    R(\theta_2,\theta_1;t).
    \label{eq:merger_rate_symmetry}
\end{align}
Environmental, halo, or orbital dependences may be included in the state or supplied as explicit conditioning variables.

Conditional on a merger, let $P_{\rm rem}(\pd\theta\mid\theta_1,\theta_2;t)$ be the remnant-state
kernel. 
For a state domain closed under mergers,
\begin{align}
    P_{\rm rem}
    \left(
        \mathcal M
        \mid
        \theta_1,\theta_2;t
    \right)
    &=1,
    \label{eq:remnant_kernel_normalization}
\end{align}
and the kernel is taken to be symmetric under exchange of the progenitors. 
If some remnants leave the modeled domain, it may instead be a sub-probability kernel, with missing mass interpreted as removal.
When it admits a density, write
\begin{align}
    P_{\rm rem}
    \left(
        \pd\theta
        \mid
        \theta_1,\theta_2;t
    \right)
    &=
    p_{\rm rem}
    \left(
        \theta
        \mid
        \theta_1,\theta_2;t
    \right)
    \pd V_{\mathcal M}.
    \notag
\end{align}

The merger contribution to the population equation is then
\begin{align}
    J_t[\rho](\theta)
    &=
    \frac{1}{2}
    \int_{\mathcal M}
    \int_{\mathcal M}
    R(\theta_1,\theta_2;t)
    p_{\rm rem}
    \left(
        \theta
        \mid
        \theta_1,\theta_2;t
    \right)
    \rho(\theta_1,t)
    \rho(\theta_2,t)
    \,\pd V_{\mathcal M,1}
    \,\pd V_{\mathcal M,2}
    \notag\\
    &\quad
    -
    \rho(\theta,t)
    \int_{\mathcal M}
    R(\theta,\theta';t)
    \rho(\theta',t)
    \,\pd V_{\mathcal M}(\theta').
    \label{eq:merger_jump_operator}
\end{align}
The first term creates remnants near $\theta$; the factor $1/2$ avoids double counting the two progenitor orderings. 
The second removes a galaxy in state $\theta$ when it merges with any partner. 
No factor $1/2$ appears there because both progenitors disappear. 
The kernels $R$ and $P_{\rm rem}$ summarize population-level encounter frequencies and conditional outcomes at the adopted coarse-graining.

\subsection{Remnant-state distributions and conservation laws}
\label{subsec:remnant_distribution_conservation}

The coarse-grained progenitor states need not determine a unique remnant. 
Let $u\in\mathcal U$ collect event-level variables, such as orbital geometry or unresolved gas structure, and write
\begin{align}
    \theta_{\rm rem}
    &=
    T(\theta_1,\theta_2,u;t),
    \notag
\end{align}
with conditional distribution $q(\pd u\mid\theta_1,\theta_2;t)$. The remnant kernel is the pushforward of this distribution:
\begin{align}
    P_{\rm rem}
    \left(
        A
        \mid
        \theta_1,\theta_2;t
    \right)
    &=
    \int_{\mathcal U}
    \mathbbm{1}_{A}
    \left[
        T(\theta_1,\theta_2,u;t)
    \right]
    q
    \left(
        \pd u
        \mid
        \theta_1,\theta_2;t
    \right)
    \label{eq:remnant_pushforward}
\end{align}
for measurable $A\subset\mathcal M$. 
Its width can therefore encode physically meaningful variables omitted from the persistent effective
state, rather than only observational scatter.

For a measurable weight $w(\theta)$, let 
\begin{align}
    \mathcal M_w(t)
    &=
    \int_{\mathcal M}w(\theta)\rho(\theta,t) \,\pd V_{\mathcal M}
    \notag
\end{align}
and define $\overline w_{\rm rem}(\theta_1,\theta_2;t)$ as its conditional expectation under $P_{\rm rem}$. 
The merger contribution is
\begin{align}
    \left.
    \frac{\pd\mathcal M_w(t)}{\pd t}
    \right|_{\rm merger}
    &=
    \frac{1}{2}
    \int_{\mathcal M}
    \int_{\mathcal M}
    R(\theta_1,\theta_2;t)
    \rho(\theta_1,t)
    \rho(\theta_2,t)
    \notag\\
    &\quad\times
    \left[
        \overline w_{\rm rem}(\theta_1,\theta_2;t)
        -w(\theta_1)-w(\theta_2)
    \right]
    \,\pd V_{\mathcal M,1}
    \,\pd V_{\mathcal M,2}.
    \label{eq:merger_moment_balance}
\end{align}
A quantity is conserved by the merger channel in conditional mean when
\begin{align}
    \overline w_{\rm rem}(\theta_1,\theta_2;t)
    &=
    w(\theta_1)+w(\theta_2).
    \label{eq:additive_moment_condition}
\end{align}
For $w=1$, the bracket in
Equation~\eqref{eq:merger_moment_balance} equals $-1$, giving
\begin{align}
    \left.
    \frac{\pd N_{\rm tot}(t)}{\pd t}
    \right|_{\rm merger}
    &=
    -\frac{1}{2}
    \int_{\mathcal M}
    \int_{\mathcal M}
    R(\theta_1,\theta_2;t)
    \rho(\theta_1,t)
    \rho(\theta_2,t)
    \,\pd V_{\mathcal M,1}
    \,\pd V_{\mathcal M,2}
    \leq0.
    \label{eq:merger_number_balance}
\end{align}
Thus every binary merger reduces the galaxy count by one even when an
additive quantity is conserved. Stellar mass may be approximately
additive, depending on stripping, star formation, and the adopted
population boundary, whereas star-formation activity, morphology, and
metallicity are generally not. Higher-order interaction channels can be
constructed analogously with multi-progenitor kernels
\citep{takeuchi2026measuretheoretic}.

\subsection{Classical limit and finite-time realization}
\label{subsec:classical_finite_merger}

If the state consists only of a non-negative additive mass $M$ and the
remnant mass is $M_1+M_2$, substitution of the corresponding Dirac
remnant kernel into Equation~\eqref{eq:merger_jump_operator} gives
\begin{align}
    J_t[\rho](M)
    &=
    \frac{1}{2}
    \int_0^M
    R(M',M-M';t)
    \rho(M',t)
    \rho(M-M',t)
    \,\pd M'
    \notag\\
    &\quad
    -
    \rho(M,t)
    \int_0^\infty
    R(M,M';t)
    \rho(M',t)
    \,\pd M'.
    \label{eq:smoluchowski_limit}
\end{align}
This is the classical Smoluchowski coagulation operator
\citep{smoluchowski1917versuch}. The multivariate theory retains the
same gain-and-loss structure while replacing deterministic mass
addition by a probability kernel on the effective state space.

At discrete epochs, each initial galaxy over an interval $[s,t]$ is
assigned to a non-merger channel, a primary channel in merger class
$c\in\mathcal C_{\rm merg}$, or a secondary channel. Let the respective
probabilities be $r^{(0)}_{t,s}(\theta)$,
$r^{\rm pri,c}_{t,s}(\theta)$, and
$r^{\rm sec}_{t,s}(\theta)$. For the restricted cohort,
\begin{align}
    r^{(0)}_{t,s}(\theta)
    +r^{\rm sec}_{t,s}(\theta)
    +\sum_{c\in\mathcal C_{\rm merg}}
    r^{\rm pri,c}_{t,s}(\theta)
    &=1.
    \label{eq:finite_merger_channel_partition}
\end{align}
The non-merger destination is described by
$\mathcal Q^{(0)}_{t,s}$ from
Section~\ref{subsec:finite_time_transition}. Conditional on being a
primary in class $c$, let
$\overline{\mathcal P}^{(c)}_{t,s}(\pd\theta'\mid\theta)$ denote the
final remnant-state kernel. It marginalizes over the partner state,
merger time, event-level variables, and post-merger evolution, and
therefore depends on the background population and tracking convention.

The reduced finite-time population kernel is
\begin{align}
    \mathcal A_{t,s}
    \left(
        A\mid\theta
    \right)
    &=
    r^{(0)}_{t,s}(\theta)
    \mathcal Q^{(0)}_{t,s}
    \left(
        A\mid\theta
    \right)+
    \sum_{c\in\mathcal C_{\rm merg}}
    r^{\rm pri,c}_{t,s}(\theta)
    \overline{\mathcal P}^{(c)}_{t,s}
    \left(
        A\mid\theta
    \right).
    \label{eq:reduced_finite_time_kernel}
\end{align}
The secondary channel has no positive destination contribution, so
\begin{align}
    \mathcal A_{t,s}
    \left(
        \mathcal M\mid\theta
    \right)
    &=
    1-r^{\rm sec}_{t,s}(\theta).
    \label{eq:reduced_kernel_total_mass}
\end{align}
For a closed cohort, propagation by this sub-probability kernel yields
\begin{align}
    \mathsf N_t^{\rm red}(A)
    &=
    \int_{\mathcal M}
    \mathcal A_{t,s}
    \left(
        A\mid\theta
    \right)
    \mathsf N_s(\pd\theta),
    \notag\\
    \mathsf N_t^{\rm red}(\mathcal M)
    &=
    \mathsf N_s(\mathcal M)
    -
    \int_{\mathcal M}
    r^{\rm sec}_{t,s}(\theta)
    \mathsf N_s(\pd\theta).
    \label{eq:reduced_finite_time_count_balance}
\end{align}
The second line is the finite-time counterpart of Equation~\eqref{eq:merger_number_balance}: when each resolved binary merger has one identified secondary, the subtracted term is the expected number of resolved events.

For a partition
\begin{align}
    \mathcal M
    &=
    \bigcup_i C_i,
    \notag
\end{align}
let $Q^{(0)}_{ji}$ be the conditional non-merger transition probability and $\overline P^{(c)}_{ji}$ the conditional primary-to-remnant transition probability. 
The reduced matrix is
\begin{align}
    A_{ji}
    &=
    r_i^{(0)}Q^{(0)}_{ji}
    +
    \sum_{c\in\mathcal C_{\rm merg}}
    r_i^{\rm pri,c}\overline P^{(c)}_{ji},
    \qquad
    \sum_j A_{ji}=1-r_i^{\rm sec}.
    \label{eq:discrete_reduced_merger_operator}
\end{align}
The predicted population is
\begin{align}
    n_j^{\rm red}(t)
    &=
    \sum_{i=1}^{N_{\rm cell}}
    A_{ji}n_i(s),
    \qquad
    n_i(s)=\mathsf N_s(C_i).
    \label{eq:discrete_reduced_merger_update}
\end{align}
Although this update is linear after calibration, the pair dependence of the merger process has been absorbed into the channel probabilities and conditional remnant kernels. 
The matrix is therefore a population-dependent finite-time realization of the quadratic merger process and must be recalibrated when the background population, interval, merger definition, or physical model changes.
Section~\ref{sec:proof_of_concept} estimates this operator from merger trees and tests it on held-out lineages.

Having specified the intrinsic transport and merger channels, we now turn to the observational forward model and state inference.

\section{Observational Forward Model and State Inference}
\label{sec:state_inference}

Sections~\ref{sec:effective_states}--\ref{sec:merger_chapter}
defined the effective state, the intrinsic population measure, and its
transport and merger dynamics. We now connect those objects to observed
galaxy catalogs. The observational kernel
$\mathcal K_{\rm obs}(\pd\bm{x}\mid\theta,t)$ was defined in
Section~\ref{subsec:population_measures}; here it supplies the likelihood,
while the evolving population supplies the epoch-dependent prior. The
distinction among measured variables $\bm{x}$, learned representations
$\bm{z}$, and effective physical states $\theta$ remains in force.

\subsection{Dynamically propagated population priors}
\label{subsec:dynamical_population_prior}

Let $\rho_0(\theta)=\rho(\theta,t_0)$ at a reference epoch, and separate
the model parameters as
$\bm\beta=(\bm\beta_{\rm dyn},\bm\beta_{\rm obs})$. For fixed dynamical
parameters, let
$\mathfrak U^{\bm\beta_{\rm dyn}}_{t,t_0}$ denote the evolution map
generated by the transport--jump model. When a density exists,
\begin{align}
    \rho_{\bm\beta_{\rm dyn}}(\theta,t;\rho_0)
    &=
    \left[
        \mathfrak U^{\bm\beta_{\rm dyn}}_{t,t_0}\rho_0
    \right](\theta).
    \label{eq:dynamically_propagated_density}
\end{align}
This notation covers both the continuous-time solution of
Equation~\eqref{eq:transport_jump_decomposition} and finite-time
implementations based on the kernels of Sections~\ref{subsec:finite_time_transition}
and~\ref{subsec:classical_finite_merger}. Because merger rates may depend
on the population itself, $\mathfrak U$ need not be linear.

The corresponding normalized state distribution is
\begin{align}
    \pi_{\bm\beta_{\rm dyn},t}(\pd\theta;\rho_0)
    &=
    \frac{
        \rho_{\bm\beta_{\rm dyn}}(\theta,t;\rho_0)
        \,\pd V_{\mathcal M}
    }{
        \displaystyle
        \int_{\mathcal M}
        \rho_{\bm\beta_{\rm dyn}}(\vartheta,t;\rho_0)
        \,\pd V_{\mathcal M}(\vartheta)
    }.
    \label{eq:dynamical_state_prior}
\end{align}
This is a population-level prior, not a deterministic trajectory for an
individual galaxy. Transition scatter, remnant-state uncertainty, and
operator uncertainty all contribute to its width. Its essential role is
to constrain populations at different epochs to arise from a common
initial population and a common evolution model.

\subsection{Likelihood for observed galaxy catalogs}
\label{subsec:catalog_likelihood}

Consider survey or redshift bins $b=1,\ldots,N_{\rm bin}$ at epochs
$t_b$, with known exposure factors $\mathcal V_b$. The latter may be an
effective comoving volume, or unity when $\mathsf N_{t_b}$ already
represents an expected catalog count. For an exact observed vector, the
predicted catalog intensity is
\begin{align}
    \lambda_b(\bm{x}\mid\rho_0,\bm\beta)
    &=
    \mathcal V_b
    \int_{\mathcal M}
    k_{{\rm obs},b}
    \left(
        \bm{x}\mid\theta,t_b;\bm\beta_{\rm obs}
    \right)
    \rho_{\bm\beta_{\rm dyn}}(\theta,t_b;\rho_0)
    \,\pd V_{\mathcal M},
    \label{eq:catalog_observational_intensity}\\
    \Lambda_b(\rho_0,\bm\beta)
    &=
    \int_X\lambda_b(\bm{x}\mid\rho_0,\bm\beta)\,\pd\bm{x}
    \notag\\
    &=
    \mathcal V_b
    \int_{\mathcal M}
    \mathcal K_{{\rm obs},b}
    \left(
        X\mid\theta,t_b;\bm\beta_{\rm obs}
    \right)
    \rho_{\bm\beta_{\rm dyn}}(\theta,t_b;\rho_0)
    \,\pd V_{\mathcal M}.
    \label{eq:catalog_expected_count}
\end{align}
Equation~\eqref{eq:catalog_observational_intensity} predicts the
distributional shape, while Equation~\eqref{eq:catalog_expected_count}
preserves the expected number of detected galaxies.

For exact catalog records
$\mathcal D_b=\{\bm{x}_{ab}\}_{a=1}^{N_b}$, a baseline model is an
inhomogeneous Poisson point process,
\begin{align}
    \mathcal L_{\rm PPP}
    (\rho_0,\bm\beta;\mathcal D_{\rm obs})
    &=
    \prod_{b=1}^{N_{\rm bin}}
    \exp[-\Lambda_b(\rho_0,\bm\beta)]
    \prod_{a=1}^{N_b}
    \lambda_b(\bm{x}_{ab}\mid\rho_0,\bm\beta).
    \label{eq:catalog_poisson_likelihood}
\end{align}
Equation~\eqref{eq:catalog_poisson_likelihood} is a likelihood for the
catalog itself rather than for a pre-estimated projected statistic.
Survey masks, measurement uncertainty, incompleteness, and censoring are
encoded in $\mathcal K_{{\rm obs},b}$. A set-valued record such as an
upper limit is handled by integrating the observational intensity over
the compatible region $E_{ab}\subset X$. Conditioning on the observed
counts instead gives a shape-only likelihood proportional to
$\prod_{ab}\lambda_b(\bm{x}_{ab})/\Lambda_b$; this intentionally discards
information about number-changing processes. Extra-Poisson variation
from clustering or cosmic variance may be added hierarchically without
changing the state-to-observation map.

\subsection{Individual and joint state-and-dynamics inference}
\label{subsec:joint_state_dynamics_inference}

Let $d_{ab}$ denote an observational record and define its object-level
factor by
\begin{align}
    \ell_{ab}(d_{ab}\mid\theta,t_b)
    &=
    \begin{cases}
        k_{{\rm obs},b}(\bm{x}_{ab}\mid\theta,t_b),
        & d_{ab}=\bm{x}_{ab},\\
        \mathcal K_{{\rm obs},b}(E_{ab}\mid\theta,t_b),
        & d_{ab}=E_{ab}.
    \end{cases}
    \notag
\end{align}
For fixed global quantities, the state posterior is
\begin{align}
    p(\theta_{ab}\mid d_{ab},t_b,\rho_0,\bm\beta)
    &=
    \frac{
        \ell_{ab}(d_{ab}\mid\theta_{ab},t_b)
        \rho_{\bm\beta_{\rm dyn}}(\theta_{ab},t_b;\rho_0)
    }{
        \displaystyle
        \int_{\mathcal M}
        \ell_{ab}(d_{ab}\mid\vartheta,t_b)
        \rho_{\bm\beta_{\rm dyn}}(\vartheta,t_b;\rho_0)
        \,\pd V_{\mathcal M}(\vartheta)
    }.
    \label{eq:individual_state_posterior}
\end{align}
In Equation~\eqref{eq:individual_state_posterior}, the observational
kernel supplies the likelihood and the propagated population supplies
the prior. Because selection is already included in
the sub-probability kernel, it must not be applied a second time.

Let $\bm\Theta_{\rm gal}=\{\theta_{ab}\}$ collect the latent states of
all catalog objects. Up to the appropriate reference measures, their
joint posterior can be written compactly as
\begin{align}
    &p(\bm\Theta_{\rm gal},\rho_0,\bm\beta
    \mid\mathcal D_{\rm obs})
    \notag\\
    &\quad\propto
    p(\rho_0,\bm\beta)
    \prod_{b=1}^{N_{\rm bin}}
    \exp[-\Lambda_b(\rho_0,\bm\beta)]
    \prod_{b=1}^{N_{\rm bin}}
    \prod_{a=1}^{N_b}
    \left[
        \mathcal V_b
        \ell_{ab}(d_{ab}\mid\theta_{ab},t_b)
        \rho_{\bm\beta_{\rm dyn}}(\theta_{ab},t_b;\rho_0)
    \right].
    \label{eq:joint_state_dynamics_posterior}
\end{align}
Equation~\eqref{eq:joint_state_dynamics_posterior} realizes the joint
inferential target stated in the Introduction. Integrating over the
latent states recovers the catalog likelihood in
Equation~\eqref{eq:catalog_poisson_likelihood}; fixing the global
quantities recovers the individual posteriors. After marginalization, galaxy-state estimates are coupled
through their shared population and evolution parameters. Multiple
redshift bins are likewise coupled through
Equation~\eqref{eq:dynamically_propagated_density}, even when individual
galaxies cannot be matched across epochs. Repeated observations or
matched descendants can be added as a longitudinal channel using the
appropriate non-merger or merger transition kernel.

\subsection{Weak, static, and dynamical population priors}
\label{subsec:prior_comparison}

The contribution of cross-epoch dynamics can be isolated by holding the observational likelihood fixed and changing only the population prior.
A weak prior imposes broad restrictions on the admissible state domain without encoding detailed population structure. 
We use the term \emph{static prior} for a population prior that is not obtained by propagating a reference population through the inferred cross-epoch
dynamics. 
Depending on the validation design, such a prior may be estimated independently at the target epoch, or it may be constructed by freezing a training population at the reference epoch. 
The frozen-static construction provides a controlled baseline that isolates the contribution of cross-epoch propagation while holding the available training population fixed.

The dynamical prior uses Equation~\eqref{eq:dynamical_state_prior}. 
It makes the stronger and falsifiable claim that the target population is reachable from the reference population under the inferred evolution model. Differences between a static and a dynamical prior therefore reflect cross-epoch propagation only when their training information, state domain, and
observational likelihood are otherwise held fixed.

For a held-out test, neither the prior nor its calibration may use the held-out measurement. 
Retaining uncertainty in the initial population and dynamical parameters gives the training-based predictive mixture
\begin{align}
    \pi_{\rm dyn,t}^{\rm pred}
    (\pd\theta\mid\mathcal D_{\rm train})
    &=
    \int
    \pi_{\bm\beta_{\rm dyn},t}
    (\pd\theta;\rho_0)
    p
    \left(
        \rho_0,
        \bm\beta_{\rm dyn}
        \mid
        \mathcal D_{\rm train}
    \right)
    \,\mathcal D\rho_0
    \,\pd\bm\beta_{\rm dyn}.
    \label{eq:predictive_dynamical_prior}
\end{align}
Equation~\eqref{eq:predictive_dynamical_prior} excludes the held-out measurement while propagating calibration uncertainty. 
Comparisons among weak, static, and dynamical priors are meaningful only when the object-level likelihood, state grid, held-out sample, and permitted training information are identical.

The controlled experiment in Section~\ref{sec:proof_of_concept} uses a plug-in version of this comparison. 
Its flat prior is the weak baseline, its static prior is the normalized initial training population left unpropagated, and its dynamical prior is obtained by propagating that same training population through an independently calibrated finite-time operator. 
We refer to the second construction as the frozen-static prior when its temporal meaning must be emphasized.

\subsection{Inference through learned representations}
\label{subsec:representation_inference}

A learned representation may compress the observations. 
For a deterministic map $\bm z=\Psi_b(\bm x)$, the induced state-to-representation kernel is the pushforward
\begin{align}
    \mathcal K_{{\rm rep},b}
    (F\mid\theta,t_b)
    &=
    \mathcal K_{{\rm obs},b}
    (\Psi_b^{-1}(F)\mid\theta,t_b),
    \qquad F\subset Z.
    \label{eq:representation_observation_kernel}
\end{align}
Equation~\eqref{eq:representation_observation_kernel} carries the
selection and measurement model into representation space. A stochastic
encoder is handled by composing its kernel with $\mathcal K_{\rm obs}$. In either case, inference remains about $\theta$;
$\bm z$ is an algorithm-dependent summary of noisy, selected data. A
discriminative calibration $q_{\rm cal}(\theta\mid\bm z,t)$ also contains
the calibration-population prior and must be corrected when transferred
to a different target population, unless the full state-to-representation
model is fitted jointly.

Adequacy of the representation is task dependent. For individual-state
inference, a useful criterion is approximate posterior sufficiency,
\begin{align}
    p(\theta\mid\bm x,t,\mathcal H)
    &\simeq
    p(\theta\mid\Psi(\bm x),t,\mathcal H),
    \label{eq:representation_posterior_sufficiency}
\end{align}
where $\mathcal H$ denotes the fixed state, population, and observation
model. Inference on the dynamics requires the analogous comparison of held-out
predictive performance. If $\Psi$ is learned from the analysis catalog,
its uncertainty and data reuse must be controlled through an independent
training sample or a joint hierarchical model.

\subsection{Identifiability relative to the observation design}
\label{subsec:state_dynamics_identifiability}

Two global models are observationally equivalent for a given catalog
design when
\begin{align}
    \lambda_b(\bm x\mid\rho_0,\bm\beta)
    &=
    \lambda_b(\bm x\mid\widetilde\rho_0,
    \widetilde{\bm\beta})
    \label{eq:observational_equivalence}
\end{align}
for almost every $\bm x$ and every observed bin. Models satisfying Equation~\eqref{eq:observational_equivalence} cannot
be distinguished using those catalogs alone without additional
assumptions. A shape-only likelihood creates further equivalences because
models with different expected counts can share the same normalized
intensity. 
Retaining counts can break some degeneracies, but does not by itself identify transport, removal, merger, and selection separately.

Identifiability is therefore a property of the complete state,
dynamics, observation model, and survey design. It can be improved by
combining epochs, retaining joint measurements, adding observables that
respond selectively to particular channels, and externally calibrating
the observational kernel. Individual-state, population, operator, and
observation-model uncertainties should remain separately reported even
when regularization selects a preferred solution.

\subsection{Posterior prediction and model checking}
\label{subsec:posterior_prediction}

For a future or held-out catalog $\mathcal D_\star$, the posterior
predictive distribution is
\begin{align}
    p(\mathcal D_\star\mid\mathcal D_{\rm obs})
    &=
    \int
    p(\mathcal D_\star\mid\rho_0,\bm\beta)
    p(\rho_0,\bm\beta\mid\mathcal D_{\rm obs})
    \,\mathcal D\rho_0\,\pd\bm\beta.
    \label{eq:posterior_predictive_catalog}
\end{align}
Each draw from Equation~\eqref{eq:posterior_predictive_catalog} first
propagates the intrinsic population to the target epoch and then applies
the target observational kernel. Useful checks include
catalog counts, joint observational distributions, detection and
censoring fractions, transition occupancies, merger-channel frequencies,
remnant-state distributions, and coverage of individual-state
posteriors. Agreement with a single luminosity or stellar-mass function
is not sufficient.

Posterior prediction also tests the state definition. Residual dependence on omitted variables, systematic failure of Equation~\eqref{eq:chapman_kolmogorov_effective}, or poor transfer across time intervals indicates that the state, temporal resolution, or operator class should be revised.

Section~\ref{sec:proof_of_concept} implements two complementary held-out checks. 
First, a finite-time operator calibrated exclusively on training root merger trees is applied to the initial population of held-out root merger trees, and the predicted later distribution and total count are compared with their realized values. 
Second, the realized held-out cohort is subjected to a known noisy and censored observation model, and the latent later population is reconstructed under flat, frozen-static, and dynamical priors. 
The second experiment conditions on the held-out cohort and its observed size and is therefore an object-level, shape-conditional special case of the catalog model in Section~\ref{subsec:catalog_likelihood}. 
The first experiment evaluates population shape and count evolution, whereas the second evaluates shape-conditional latent-state reconstruction for a fixed held-out cohort.

The next section examines projected population summaries as analytic examples of non-closure and information loss.

\section{Projected Observables and Astrophysical Applications}
\label{sec:applications}

Section~\ref{sec:state_inference} formulated inference at the level of
the full observational catalog. Reduced statistics such as luminosity
functions, stellar-mass functions, and quenched fractions remain useful,
but they are projections of the same generative model rather than
independent dynamical variables. This section uses a small set of
analytic examples to show what such projections retain, why they are
usually non-closed, and how additional observables can reduce the
resulting degeneracies.

\subsection{Projected balances and luminosity functions}
\label{subsec:projected_population_balance}

Let $y=g(\theta,t)$ be a differentiable scalar summary of the effective
state. Its intrinsic projected intensity is
\begin{align}
    \phi_g(y,t)
    &=
    \int_{\mathcal M}
    \rho(\theta,t)
    \deltadir\!\left[y-g(\theta,t)\right]
    \,\pd V_{\mathcal M}.
    \notag
\end{align}
Along the continuous flow, the state-level rate of change of the summary
is
\begin{align}
    a_g(\theta,t)
    &=
    \frac{\partial g}{\partial t}
    +
    \bm v\cdot\nabla_{\mathcal M}g,
    \notag
\end{align}
and the projected transport flux is
\begin{align}
    F_g(y,t)
    &=
    \int_{\mathcal M}
    \rho(\theta,t)a_g(\theta,t)
    \deltadir\!\left[y-g(\theta,t)\right]
    \,\pd V_{\mathcal M}.
    \notag
\end{align}
Projecting Equation~\eqref{eq:transport_jump_decomposition} gives the
exact balance
\begin{align}
    \frac{\partial\phi_g}{\partial t}
    +
    \frac{\partial F_g}{\partial y}
    &=
    B_g-D_g+J_g[\rho],
    \label{eq:projected_population_balance}
\end{align}
where $B_g$, $D_g$, and $J_g[\rho]$ are the corresponding projections of
$B$, $\Lambda\rho$, and $J_t[\rho]$. Where $\phi_g>0$, one may write
$F_g=A_g\phi_g$, with
\begin{align}
    A_g(y,t)
    &=
    \mathbbm E_\rho
    \left[
        a_g(\theta,t)
        \mathrel{\big|}
        g(\theta,t)=y
    \right].
    \notag
\end{align}
Equation~\eqref{eq:projected_population_balance} is therefore exact but
not generally autonomous. The conditional drift, removal, and merger
terms depend on how the population is distributed along the hidden
directions of each level set $g(\theta,t)=y$. Two state-space populations
can have the same $\phi_g$ and different projected derivatives. The same
construction applies to a vector of summaries, for which the joint
projection retains correlations but remains non-closed whenever distinct
effective states share the same projected value and have different
transition laws.

The intrinsic projection must also be distinguished from a statistic
measured by a survey. If $h_b:X\rightarrow Y_b$ is a summary of the
recorded data in bin $b$, its expected observed measure is
\begin{align}
    \mathsf M_{h,b}(E)
    &=
    \mathcal V_b
    \int_{\mathcal M}
    \mathcal K_{{\rm obs},b}
    \left(
        h_b^{-1}(E)
        \mid
        \theta,t_b
    \right)
    \rho(\theta,t_b)
    \,\pd V_{\mathcal M}.
    \notag
\end{align}
Thus, a measured luminosity or mass function is a marginal of the
catalog model and generally includes selection, noise, and censoring. It
coincides with the intrinsic pushforward only in an idealized complete
and error-free observation process.

For an intrinsic luminosity $L=\ell(\theta,t)>0$, the projected balance
becomes
\begin{align}
    \phi_L(L,t)
    &=
    \int_{\mathcal M}
    \rho(\theta,t)
    \deltadir\!\left[L-\ell(\theta,t)\right]
    \,\pd V_{\mathcal M},
    \notag\\
    \frac{\partial\phi_L}{\partial t}
    +
    \frac{\partial}{\partial L}
    \left(A_L\phi_L\right)
    &=
    B_L-D_L+J_L[\rho].
    \notag
\end{align}
The merger term still depends on the full progenitor states and on the
remnant-state kernel. It closes as a one-dimensional Smoluchowski
operator only under strong additional assumptions, including a rate
kernel depending solely on the progenitor luminosities and a
deterministically additive remnant luminosity. This controlled limit
recovers the classical coagulation form used in early merger models
\citep{1978ApJ...220..390S,1978ApJ...223L..59S}, but luminosity is not in
general additive because of stellar-population evolution, dust, mass
loss, and merger-triggered star formation.

\subsection{The Schechter family as a projected invariant family}
\label{subsec:schechter_invariant_family}

The Schechter intensity density is
\begin{align}
    \phi_{\rm S}(L;\bm\eta)
    &=
    \frac{\phi_\ast}{L_\ast}
    \left(\frac{L}{L_\ast}\right)^\alpha
    \exp\!\left(-\frac{L}{L_\ast}\right),
    \qquad
    \bm\eta=
    \left(\ln\phi_\ast,\ln L_\ast,\alpha\right),
    \notag
\end{align}
following \cite{1976ApJ...203..297S}. Let
\begin{align}
    \mathcal H_L[\rho]
    &=
    -\frac{\partial}{\partial L}
    \left(A_L\phi_L\right)
    +B_L-D_L+J_L[\rho]
    \notag
\end{align}
be the luminosity-space vector field evaluated along the full
state-space solution. If $\phi_L=\phi_{\rm S}(L;\bm\eta)$ at an epoch,
the Schechter family is invariant along that trajectory only if
\begin{align}
    \mathcal H_L[\rho]
    &=
    \dot{\ln\phi_\ast}
    \frac{\partial\phi_{\rm S}}{\partial\ln\phi_\ast}
    +
    \dot{\ln L_\ast}
    \frac{\partial\phi_{\rm S}}{\partial\ln L_\ast}
    +
    \dot\alpha
    \frac{\partial\phi_{\rm S}}{\partial\alpha}.
    \notag
\end{align}
After division by $\phi_{\rm S}$, the three tangent directions are
$1$, $L/L_\ast-(1+\alpha)$, and $\ln(L/L_\ast)$. Components of
$\mathcal H_L[\rho]$ normal to this tangent space quantify departures
from a Schechter description.

The full $\rho$ remains essential. Equal Schechter parameters do not
imply equal conditional drifts, source and removal terms, or remnant
distributions, so the subsequent evolution may differ. Observed
parameter changes constrain only projected combinations. A good fit
therefore identifies an approximate low-dimensional direction in
function space, not a unique physical mechanism or a closed luminosity
model.

\subsection{Downsizing as flux and residence time}
\label{subsec:downsizing_flux_residence}

The term \emph{downsizing} describes several related mass-dependent
trends in galaxy evolution
\citep{1996AJ....112..839C,2006MNRAS.372..933N,
2010ApJ...721..193P}. Here we focus on the tendency for more massive
galaxies to leave the actively star-forming phase earlier, while
recognizing that later dry mergers can alter the assembly time of the
final stellar mass.

Let $m=m(\theta,t)$ be a stellar-mass coordinate and partition the state
space into star-forming and quiescent regions,
$\Omega_{\rm SF}(t)$ and $\Omega_{\rm Q}(t)$. The mass-resolved
star-forming intensity is
\begin{align}
    \phi_{\rm SF}(m,t)
    &=
    \int_{\Omega_{\rm SF}(t)}
    \rho(\theta,t)
    \deltadir\!\left[m-m(\theta,t)\right]
    \,\pd V_{\mathcal M}.
    \notag
\end{align}
Writing $F_{m,{\rm SF}}$ for transport along the mass coordinate and
$C_{{\rm SF}\rightarrow{\rm Q}}^{\rm tr}$ and
$C_{{\rm Q}\rightarrow{\rm SF}}^{\rm tr}$ for continuous crossing of
the moving phase boundary, the projected balance is
\begin{align}
    \frac{\partial\phi_{\rm SF}}{\partial t}
    +
    \frac{\partial F_{m,{\rm SF}}}{\partial m}
    &=
    B_{\rm SF}-D_{\rm SF}
    -C_{{\rm SF}\rightarrow{\rm Q}}^{\rm tr}
    +C_{{\rm Q}\rightarrow{\rm SF}}^{\rm tr}
    +J_{\rm SF}[\rho].
    \label{eq:star_forming_population_balance}
\end{align}
The crossing terms are defined using the galaxy velocity relative to the
classification boundary, so a changing threshold is not confused with
physical transport. The merger term remains separate because a merger
can remove two progenitors, produce one remnant, and change both phase
membership and tracked identity.

Equation~\eqref{eq:star_forming_population_balance} shows why a single
condition on the mass-direction drift is not a general explanation of
downsizing. The decline of the high-mass star-forming population can
reflect boundary crossing, merger redistribution, source or removal,
rejuvenation, or the initial population. A complementary trajectory-level
diagnostic is the first exit time
\begin{align}
    \tau_{\rm SF}
    &=
    \inf\left\{
        u>0:
        \theta_{t_0+u}\notin\Omega_{\rm SF}(t_0+u)
    \right\},
    \notag\\
    \overline\tau_{\rm SF}(m,t_0)
    &=
    \mathbbm E
    \left[
        \tau_{\rm SF}
        \mathrel{\big|}
        m(\theta_{t_0},t_0)=m,
        \theta_{t_0}\in\Omega_{\rm SF}(t_0)
    \right].
    \notag
\end{align}
Secondary-progenitor disappearance or other exits may be represented by
an absorbing state. One operational residence-time form of downsizing is
\begin{align}
    \frac{\partial}{\partial m}
    \overline\tau_{\rm SF}(m,t_0)
    &<0,
    \notag
\end{align}
over a stated mass, epoch, and cohort range. Conditions involving only a
projected velocity are recovered only in restrictive deterministic
limits with negligible jumps, rejuvenation, source, and removal terms.

\subsection{Multiple observables and identifiability}
\label{subsec:multiobservable_identifiability}

Combining observables can reduce projection degeneracies, but the gain
must be assessed through their joint catalog intensity rather than by
counting separate plotted marginals. For a joint measured vector
$\bm y$ and model parameters $\bm\beta$, define the observational
equivalence class
\begin{align}
    [\bm\beta]_{\mathcal O}
    &=
    \left\{
        \widetilde{\bm\beta}:
        \lambda_b(\bm y\mid\widetilde{\bm\beta})
        =
        \lambda_b(\bm y\mid\bm\beta)
        \ \text{for almost every $\bm y$ and every $b$}
    \right\}.
    \notag
\end{align}
An additional observable refines this class only if its conditional
distribution responds to parameter directions that were previously
invisible. Locally, this can be diagnosed by the Poisson information
matrix
\begin{align}
    \mathcal I_{\mu\nu}(\bm\beta)
    &=
    \sum_b
    \int
    \frac{
        \partial_{\beta_\mu}\lambda_b(\bm y\mid\bm\beta)
        \partial_{\beta_\nu}\lambda_b(\bm y\mid\bm\beta)
    }{
        \lambda_b(\bm y\mid\bm\beta)
    }
    \,\pd\bm y.
    \notag
\end{align}
New measurements are informative only when they add sensitivity
directions. Counts across epochs probe channels that change galaxy
number. Joint mass--SFR data probe transport across the star-forming
boundary. Merger indicators and remnant data constrain the rate and
remnant kernels, while environmental or gas-related data can expose
missing state variables.

Several marginals estimated from the same galaxies are correlated and
should not be multiplied as independent likelihoods. Conversely, fitting
only separate marginals discards cross-correlations that may distinguish
physical mechanisms. The appropriate object is the joint catalog
likelihood of Section~\ref{subsec:catalog_likelihood}, or a calibrated
approximation to it. Finally, if an added observable retains predictive
information about future transitions after conditioning on $\theta$, the
predictive-adequacy criterion calls for enlarging the effective state
rather than treating that observable merely as another projection.

These examples give a concise set of diagnostics: closure of a projected
balance, tangency to the Schechter family, mass-dependent phase flux and
residence time, and sensitivity of the joint observational intensity.
Section~\ref{sec:calibration} develops the parameterizations and
computational procedures needed to estimate these quantities.

\section{Calibration, Regularization, and Computational Inference}
\label{sec:calibration}

Sections~\ref{sec:state_inference} and~\ref{sec:applications} defined the
inferential target and showed why projected observables do not generally
identify the underlying state-space dynamics. The remaining task is to
turn the transport--jump model into an estimable generative model.
Calibration, regularization, and computation have different roles.
Positivity, normalization, progenitor-exchange symmetry, and population
count balance are structural requirements and should be enforced by
construction. Smoothness, locality, sparsity, and similarity to a
reference model are uncertain assumptions and should enter through
explicit priors whose influence can be tested. Regularization can select
a representative from an observational equivalence class, but it does
not by itself make that representative identifiable.

The continuous-time formulation admits constrained parameterizations of its tangent transport field, non-negative source intensities and removal rates, a symmetric merger-rate kernel, and a normalized remnant kernel. 
For data available only at discrete epochs, however, it is often more direct to parameterize and calibrate the finite-time transition model itself. 
The proof of concept in Section~\ref{sec:proof_of_concept} follows this finite-time route. 
It estimates the conditional non-merger transition on a relatively fine state grid, while the rarer merger-channel probabilities and primary-to-remnant distributions are estimated on a coarser source partition. 
The resulting sub-stochastic operator is then evaluated on held-out root merger trees. 
The two resolutions implement a common effective state while pooling the sparse merger statistics over coarser source bins.

\subsection{Constrained finite-time and observation models}
\label{subsec:finite_time_calibration}

Partition the effective state space into cells $C_i$ and let 
\begin{align}
    h
    &\in
    \mathcal C_{\rm ch}
    =
    \{0,{\rm sec},({\rm pri},c)\}
    \notag
\end{align}
denote the mutually
exclusive non-merger, secondary, and primary-merger channels introduced
in Section~\ref{subsec:classical_finite_merger}. Channel probabilities
and conditional destination kernels can be placed on probability
simplices, for example through
\begin{align}
    r_i^{(h)}
    =
    \frac{\exp\eta_i^{(h)}}{\sum_{h'\in\mathcal C_{\rm ch}}\exp\eta_i^{(h')}},
    \qquad 
    Q_{ji}^{(0)}
    =
    \frac{\exp q_{ji}}{\sum_k\exp q_{ki}},
    \qquad
    \overline P_{ji}^{(c)}
    =
    \frac{\exp p_{ji}^{(c)}}{\sum_k\exp p_{ki}^{(c)}}.
    \notag
\end{align}
The reduced operator is then
\begin{align}
    A_{ji}
    &=
    r_i^{(0)}Q_{ji}^{(0)}
    +
    \sum_c
    r_i^{({\rm pri},c)}
    \overline P_{ji}^{(c)},
    \notag\\
    \sum_jA_{ji}
    &=
    1-r_i^{\rm sec}.
    \label{eq:constrained_finite_time_operator}
\end{align}
The source partition used for the channel probabilities need not coincide with the state grid used for the non-merger transition and later population counts. If a fine cell $C_i$ belongs to a coarser source bin $B_{b(i)}$, one may tie the rare-channel parameters across that bin by replacing $r_i^{(h)}$ and $\overline P_{ji}^{(c)}$ with $r_{b(i)}^{(h)}$ and $\overline P_{j b(i)}^{(c)}$. 
The reduced operator then retains the same column-mass identity, 
\begin{align}
    \sum_j A_{ji}
    &=
    1-r_{b(i)}^{\rm sec}
    \notag
\end{align}
while allowing the non-merger kernel to be calibrated at a finer state-space resolution. 
This is the sparse-merger construction used in Section~\ref{sec:proof_of_concept}.
Equation~\eqref{eq:constrained_finite_time_operator} preserves non-negativity, normalization, and the loss of secondary progenitors before fitting. 
Sources or additional removal channels can be added explicitly rather than hidden in a renormalized transition matrix. Over several epochs a general update is
\begin{align}
    \bm n_{b+1}
    &=
    \mathsf A_b
    \left(
        \bm n_b;
        \bm\beta_{\rm dyn}
    \right)
    \bm n_b
    +
    \bm b_b,
    \qquad
    \bm b_b\geq\bm0.
    \notag
\end{align}
The dependence of $\mathsf A_b$ on $\bm n_b$ retains the possibility
that merger probabilities depend on the partner population. Treating a
calibrated matrix as fixed is a local approximation whose domain of
validity must be stated.

Using non-negative state basis functions $\varphi_i$, the observation
kernel defines a matrix
\begin{align}
    \mathsf O_{\alpha i}^{(b)}
    &=
    \mathcal V_b
    \int_{\mathcal M}
    \mathcal K_{{\rm obs},b}
    \left(
        E_{b\alpha}
        \mid
        \theta,t_b
    \right)
    \varphi_i(\theta)
    \,\pd V_{\mathcal M},
    \notag\\
    \lambda_{b\alpha}
    &=
    \sum_i
    \mathsf O_{\alpha i}^{(b)}n_{bi},
    \qquad
    Y_{b\alpha}\sim{\rm Poisson}(\lambda_{b\alpha}).
    \label{eq:discrete_observational_forward_model}
\end{align}
Equation~\eqref{eq:discrete_observational_forward_model} gives the binned form of the catalog likelihood introduced in Section~\ref{subsec:catalog_likelihood}.
Selection and measurement scatter remain inside the observation operator. 
In continuous or discrete implementations, the observational kernel may be factorized into a detection probability and a normalized measurement model, but its calibration should remain distinct from the intrinsic dynamics.

\subsection{Regularization and simulation-assisted calibration}
\label{subsec:simulation_assisted_calibration}

A Bayesian implementation can organize regularization through
hyperparameters $\bm\lambda$,
\begin{align}
    p
    \left(
        \rho_0,
        \bm\beta,
        \bm\lambda
        \mid
        \mathcal D
    \right)
    &\propto
    \mathcal L(\mathcal D\mid\rho_0,\bm\beta)
    p(\rho_0\mid\bm\lambda_0)
    p(\bm\beta_{\rm dyn}\mid\bm\lambda_{\rm dyn})
    p(\bm\beta_{\rm obs}\mid\bm\lambda_{\rm obs})
    p(\bm\lambda).
    \notag
\end{align}
Examples include metric-aware smoothness of the transport field,
neighboring-cell shrinkage of transition statistics, and broad reference
priors for poorly sampled remnant kernels. These assumptions should not
replace exact constraints that can be imposed structurally, and they
should not force merger remnants or long-interval transitions to remain
local. Hyperparameters must be selected without using the final held-out
sample, and sensitivity to plausible alternatives is part of the
scientific result.

Simulations supply resolved states, descendant links, merger identities, and synthetic observations that are unavailable in a survey catalog.
Here simulations calibrate and test a reduced model whose effective state is specified independently through its physical interpretation and prediction task.
For a trajectory segment $n$ spanning the epoch interval $[t_{b_n},t_{b_n+1}]$, let $i_n$ denote its initial state cell and let $h_n$ denote its channel. 
Whenever the segment has an independent final galaxy, let $j_n$ denote its destination cell. 
We use the interval-indexed notation $r_{b,i}^{(h)}$, $Q_{b,ji}^{(0)}$, and $\overline P_{b,ji}^{(c)}$ for the channel probabilities, conditional non-merger transitions, and conditional primary-to-remnant transitions, respectively. 
The finite-time calibration likelihood is
\begin{align}
    \mathcal L_{\rm sim}^{\rm tr}
    &=
    \prod_n
    r_{b_n,i_n}^{(h_n)}
    \prod_{n:\,h_n=0}
    Q_{b_n,j_n i_n}^{(0)}
    \prod_c
    \prod_{n:\,h_n=({\rm pri},c)}
    \overline P_{b_n,j_n i_n}^{(c)}.
    \label{eq:simulation_transition_likelihood}
\end{align}
Equation~\eqref{eq:simulation_transition_likelihood} separates channel
occurrence from the conditional destination distributions. A secondary
segment contributes through the channel factor
$r_{b_n,i_n}^{({\rm sec})}$ but has no destination factor because it does not survive as an independent later galaxy. 
When only one epoch interval is calibrated, as in the controlled proof of concept in Section~\ref{sec:proof_of_concept}, the interval index $b$ may be suppressed. Because segments belonging to the same merger tree are correlated, training, validation, and test partitions should be made at the level of lineages or independent realizations rather than by
randomly splitting individual transitions.

Simulations also make predictive adequacy of the state testable. If $\eta_n$
is omitted from $\theta_n$, a held-out comparison of models with and
without $\eta_n$ can use
\begin{align}
    \Delta_\eta^{\rm pred}
    &=
    \frac{1}{N_{\rm test}}
    \sum_{n\in\mathcal I_{\rm test}}
    \log
    \frac{
        p_1(Y_n\mid\theta_n,\eta_n,t_n)
    }{
        p_0(Y_n\mid\theta_n,t_n)
    }.
    \label{eq:heldout_predictive_state_gain}
\end{align}
A reproducible positive gain indicates that the proposed state is not
predictively adequate at the chosen time resolution. The state should
then be enlarged, or the operator explicitly conditioned on $\eta$,
rather than the residual variation being labeled as measurement noise.
Observation-model calibration should be performed separately with
injection tests, repeat measurements, or mock observations, and its
uncertainty propagated into the final posterior.

\subsection{Transfer and posterior computation}
\label{subsec:likelihood_and_sbi}

A simulator-domain operator cannot in general be transferred unchanged
to a survey population. Resolution, subgrid prescriptions, cosmology,
sample definition, and omitted environment variables may change both
channel frequencies and destination distributions. A survey analysis
should therefore distinguish simulator and target operators, introduce
explicit discrepancy priors, and report regions where the target lies
outside the simulation support. Selection, blending, noise, and
censoring must in any case be calibrated for the target survey.

Let $\bm\vartheta$ collect the initial population, dynamical,
observation, transfer, and state-map parameters. The joint posterior can
be written schematically as
\begin{align}
    p
    \left(
        \bm\Theta_{\rm gal},
        \bm\vartheta
        \mid
        \mathcal D_{\rm obs},
        \mathcal D_{\rm sim}^{\rm tr},
        \mathcal D_{\rm cal}^{\rm obs}
    \right)
    &\propto
    \mathcal L_{\rm obs}
    \mathcal L_{\rm sim}^{\rm tr}
    \mathcal L_{\rm cal}^{\rm obs}
    p(\bm\vartheta).
    \label{eq:joint_observation_simulation_posterior}
\end{align}
Equation~\eqref{eq:joint_observation_simulation_posterior} combines the survey likelihood, simulation-transition calibration, and observation-model calibration as distinct information sources.
When these factors are tractable, ordinary likelihood-based computation is sufficient. 
A modular analysis may instead hold externally calibrated operator and observation modules fixed while testing them on a target catalog, provided that the resulting cut in posterior feedback is stated explicitly.

When the likelihood cannot be evaluated, a forward simulator can be
used after the generative model has been specified:
\begin{align}
    \bm\vartheta^{(m)}
    &\sim
    p_{\rm train}(\bm\vartheta), \notag 
    \\
        \mathcal D_{\rm syn}^{(m)}
    &\sim
    p_{\rm fw}
    \left(
        \,\cdot\,
        \mid
        \bm\vartheta^{(m)}
    \right), \notag 
    \\
    q_\phi
    \left(
        \bm\vartheta
        \mid
        \mathsf s_{\rm cat}(\mathcal D_{\rm obs})
    \right)
    &\simeq
    p
    \left(
        \bm\vartheta
        \mid
        \mathsf s_{\rm cat}(\mathcal D_{\rm obs})
    \right).
    \label{eq:sbi_posterior_target}
\end{align}
Equation~\eqref{eq:sbi_posterior_target} illustrates neural posterior
estimation; related methods may approximate a likelihood or likelihood ratio \citep{2019MNRAS.488.4440A,2020PNAS..11730055C}. 
SBI provides posterior computation for the specified transport--jump and observation model, using catalog summaries or neural encoders as algorithm-dependent representations of the observations.
The training distribution must cover the posterior region, and the forward simulator must include the same selection, noise, censoring, operator uncertainty, and transfer assumptions as the scientific model.

\subsection{Uncertainty propagation and validation}
\label{subsec:uncertainty_calibration_validation}

A point estimate of the operator is insufficient. Its posterior
uncertainty should be propagated according to
\begin{align}
    p
    \left(
        \bm n_{b+1}
        \mid
        \bm n_b,
        \mathcal D_{\rm sim}^{\rm tr}
    \right)
    &=
    \int
    p
    \left(
        \bm n_{b+1}
        \mid
        \bm n_b,
        \bm\beta_{\rm dyn}
    \right)
    p
    \left(
        \bm\beta_{\rm dyn}
        \mid
        \mathcal D_{\rm sim}^{\rm tr}
    \right)
    \,\pd\bm\beta_{\rm dyn}.
    \label{eq:operator_uncertainty_propagation}
\end{align}
Equation~\eqref{eq:operator_uncertainty_propagation} separates stochastic transition variation from uncertainty due to finite calibration data, regularization, and model transfer.

Validation must distinguish computational calibration from scientific adequacy. 
Coverage tests and simulation-based calibration assess computational correctness under the assumed model, whereas scientific adequacy requires validation on held-out lineages, epochs, or survey regions.
Relevant checks include proper transition scores, channel probabilities, destination distributions, total counts, secondary-progenitor number loss, merger frequencies, remnant states, posterior coverage, the count balance in Equation~\eqref{eq:reduced_finite_time_count_balance}, and the Chapman--Kolmogorov relation in Equation~\eqref{eq:chapman_kolmogorov_effective}. Residual dependence on
omitted variables returns to
Equation~\eqref{eq:heldout_predictive_state_gain}.

Section~\ref{sec:proof_of_concept} implements a restricted plug-in
version of this machinery. It fixes an interpretable two-dimensional
effective physical state and divides the sample at the level of root
merger trees. Training root merger trees are used to estimate a finite-time
non-merger kernel and a reduced finite-time operator with explicit merger channels, which are then tested
on the initial population of held-out root merger trees. The forward comparison
distinguishes an unchanged static baseline, a count-preserving
non-merger-only counterfactual, and the full transport-plus-jump
prediction.

The subsequent state-recovery experiment conditions on the held-out cohort and combines a Gaussian measurement model with censored sSFR records. 
It compares a flat prior, the frozen initial training distribution, and the same training distribution propagated through the calibrated operator. 

\section{A Controlled IllustrisTNG Proof of Concept}
\label{sec:proof_of_concept}

We use IllustrisTNG as a controlled environment in which galaxy states, descendant links, and resolved merger histories are available \citep{2018MNRAS.473.4077P,2019ComAC...6....2N}. 
We test a restricted finite-time realization of the framework with a fixed, interpretable state and a known observation model. 
We estimate a non-merger kernel and a reduced finite-time operator with explicit merger channels from training root merger trees, predict the later population of held-out root merger trees, and use the propagated population as a prior for latent-state recovery.
We estimate a non-merger kernel and a reduced finite-time operator with explicit merger channels from training root merger trees, predict the later population of held-out root merger trees, and use the propagated population as a prior for latent-state recovery.

\subsection{State, sample, and finite-time operator}
\label{subsec:tng_state_sample}

We analyze the TNG100-1 realization of the IllustrisTNG suite at its standard
resolution, using snapshots 72 ($z=0.400$, cosmic time
$t_0=9.39\,\mathrm{Gyr}$) and 84 ($z=0.197$,
$t_1=11.32\,\mathrm{Gyr}$), so that
$\Delta t=t_1-t_0=1.93\,\mathrm{Gyr}$. The initial-snapshot selection is
the physical stellar-mass cut $M_\ast(t_0)>10^{9}\,M_\odot$. The total
bound stellar mass is obtained from the particle-type-4 (stellar) component of
\texttt{SubhaloMassType} and converted from $10^{10}M_\odot/h$ to physical
solar masses using $h=0.6774$. We use \texttt{SubhaloSFR}, the total
instantaneous SFR of gravitationally bound gas cells, without an additional
aperture correction. No \texttt{SubhaloFlag} quality cut is imposed. The
parent cohort consists of galaxies that satisfy the initial-snapshot
selection and can be tracked reliably to the later snapshot. For each retained
galaxy, the effective physical state is
\begin{align}
    \theta
    &=
    (m,s),
    \notag\\
    m
    &=
    \log_{10}
    \left(
        \frac{M_\ast}{M_\odot}
    \right),
    \notag\\
    s
    &=
    \begin{cases}
        \log_{10}
        \left(
            \dfrac{\mathrm{sSFR}}{\mathrm{yr}^{-1}}
        \right),
        &
        \mathrm{SFR}>0,
        \\[4pt]
        s_{\rm zero},
        &
        \mathrm{SFR}\leq0,
    \end{cases}
    \qquad
    s_{\rm zero}=-13.
    \label{eq:tng_effective_state}
\end{align}
The zero-SFR flag is retained separately, while $s_{\rm zero}$ supplies a finite coding value for systems with zero instantaneous SFR. 
Galaxies with resolved but very small positive SFR retain their true $\log_{10}(\mathrm{sSFR}/\mathrm{yr}^{-1})$ values even when these lie below $s_{\rm zero}$; this occurs for 271 of the 19137 galaxies ($1.4\%$) in the present sample. The quenched classification used for the
coarse source partition and the quenched-fraction diagnostic is the fixed
threshold $s\leq-11.2$. The same numerical value is used separately as the
mock-observation censoring threshold $s_{\rm lim}$ in
Section~\ref{subsec:tng_state_recovery}.

Merger-tree information assigns each initial galaxy to one of three mutually
exclusive channels: a non-merger channel with one retained descendant, a
primary channel associated with a retained merger remnant, or a secondary
channel that does not survive as an independent later galaxy. A resolved
binary event has one primary progenitor and one secondary progenitor and
satisfies
\begin{align}
    q_{\rm merg}
    &=
    \frac{M_{\ast,\rm sec}(t_0)}{M_{\ast,\rm pri}(t_0)}
    \geq0.1.
    \notag
\end{align}
Here, ``primary'' denotes the SubLink main-branch progenitor identified by
\texttt{FirstProgenitorID}, while the secondary is the sibling progenitor
reached through \texttt{NextProgenitorID}. Thus, the labels are defined by the
SubLink branch convention rather than by the stellar-mass ordering at $t_0$.
Consequently, $q_{\rm merg}$ is not restricted to values below unity: 30 of
the 520 resolved events ($5.8\%$) have $q_{\rm merg}>1$. This
interval-start mass ratio is an operational convention for the present
one-step experiment; it is evaluated from the progenitor masses at $t_0$,
not at the merger epoch or at peak historical mass. Histories that cannot be
represented as one resolved two-to-one event, including multiple or ambiguous
mergers and missing descendants, are excluded from the main cohort and
recorded in the extraction diagnostics.

For the full retained cohort before the tree-level split, the event accounting
is
\begin{align}
    N_0
    &=
    19137
    =
    18097_{\rm nonmerger}
    +
    520_{\rm primary}
    +
    520_{\rm secondary},
    \notag\\
    N_1
    &=
    18617
    =
    18097_{\rm nonmerger}
    +
    520_{\rm remnant},
    \notag\\
    N_1
    &=
    N_0-N_{\rm secondary}
    =
    19137-520
    =
    18617.
    \label{eq:tng_count_loss}
\end{align}
This identity is the empirical counterpart of
Equation~\eqref{eq:reduced_finite_time_count_balance}: each resolved binary
event contributes two initial progenitors and one later remnant.

The sample is split $70\%$--$30\%$ at the level of root merger trees, so
progenitors and remnants from the same lineage cannot enter different
subsets. The first eight hexadecimal digits of a stable SHA1 hash formed from seed 42
and the root-tree identifier are mapped to $[0,1)$; values below $0.30$ are
assigned to the held-out subset.
This gives 13413 initial galaxies in the training subset and 5724 in the
held-out subset. The $(m,s)$ domain is first partitioned into a rectangular
$10\times10$ grid. Source cells containing fewer than 30 training non-merger
galaxies are combined with neighboring cells, with an $8\times8$ grid used
only as a fallback if the resulting partition remains too sparse. The fallback
is not triggered here. The initial grid has 10 equal-width mass bins spanning
$[8.93,12.36]$ and 10 equal-width sSFR bins spanning $[-14.56,-8.29]$,
with both ranges padded by $2\%$ beyond the training range. The 100 initial
rectangular cells are merged into an adaptive partition of
$N_{\rm cell}=44$ cells, and both $\widehat{\mathsf Q}^{(0)}$ and
$\widehat{\mathsf A}$ act on this $44$-cell partition. For visualization,
cellwise quantities on the adaptive partition are mapped back to the
constituent cells of the original $10\times10$ plotting grid.

A coarser source partition $B_b$ is used for the merger-channel frequencies
and remnant distributions. Five stellar-mass edges,
$9.00$, $9.28$, $9.63$, $10.15$, and $12.30$ in
$\log_{10}(M_\ast/M_\odot)$, define four initial mass bins, which are crossed
with the division into quenched ($s\leq-11.2$) and star-forming
($s>-11.2$) systems. After adjacent mass bins containing fewer than 20
primary events are merged, the quenched side forms one mass bin and the
star-forming side retains three, giving $N_{\rm bin}=4$ realized coarse bins.
Let $b(i)$ denote the coarse bin containing adaptive cell $C_i$. States
outside the fitted rectangular domain are assigned to the nearest boundary
cell and are not excluded.

Let $N_{ji}^{(0)}$ be the number of training non-merger galaxies that move from
$C_i$ to $C_j$, and let
\begin{align}
    N_i^{(0)}
    &=
    \sum_jN_{ji}^{(0)}.
    \notag
\end{align}
The conditional finite-time non-merger transition matrix is estimated with the
fixed pseudocount $\alpha=0.5$,
\begin{align}
    \widehat Q_{ji}^{(0)}
    &=
    \frac{
        N_{ji}^{(0)}+\alpha
    }{
        N_i^{(0)}+\alpha N_{\rm cell}
    },
    \qquad
    \sum_j\widehat Q_{ji}^{(0)}=1.
    \label{eq:tng_nonmerger_operator}
\end{align}
This is a finite-time one-galaxy transition kernel, not an estimate of an
instantaneous velocity field. The mean displacements shown below are used only
for visualization; all forward predictions use the $44\times44$ matrix
$\widehat{\mathsf Q}^{(0)}=(\widehat Q_{ji}^{(0)})_{j,i}$.

For the coarse source bin $B_b$, let $N_b^{(h)}$ be the number of initial
training galaxies in channel $h\in\{0,{\rm pri},{\rm sec}\}$, and let
$N_b=\sum_hN_b^{(h)}$. The channel probabilities are
\begin{align}
    \widehat r_b^{(h)}
    &=
    \frac{
        N_b^{(h)}+\alpha
    }{
        N_b+3\alpha
    },
    \qquad
    \sum_h\widehat r_b^{(h)}=1.
    \label{eq:tng_channel_probabilities}
\end{align}
Let $N_{jb}^{\rm pri}$ count primary progenitors in $B_b$ whose remnants lie
in $C_j$. After merging adjacent coarse mass bins whenever a bin contains
fewer than 20 primary events, the conditional primary-to-remnant distribution
is
\begin{align}
    \widehat{\overline P}_{jb}^{\rm pri}
    &=
    \frac{N_{jb}^{\rm pri}}{N_b^{\rm pri}},
    \qquad
    \sum_j\widehat{\overline P}_{jb}^{\rm pri}=1.
    \notag
\end{align}
The reduced finite-time operator is then
\begin{align}
    \widehat A_{ji}
    &=
    \widehat r_{b(i)}^{(0)}
    \widehat Q_{ji}^{(0)}
    +
    \widehat r_{b(i)}^{\rm pri}
    \widehat{\overline P}_{j b(i)}^{\rm pri},
    \notag\\
    \widehat{\mathsf A}
    &=
    \left(
        \widehat A_{ji}
    \right)_{j,i},
    \notag\\
    \sum_j\widehat A_{ji}
    &=
    1-\widehat r_{b(i)}^{\rm sec}.
    \label{eq:tng_reduced_operator}
\end{align}
Thus, $\widehat{\mathsf A}$ is sub-stochastic. Its missing column mass is the estimated probability that an initial galaxy is a secondary progenitor and therefore has no independent later descendant. 
The operator is a population-dependent finite-time reduction of the underlying quadratic transport--jump dynamics.

\begin{figure}[t]
    \centering
    \includegraphics[width=0.98\linewidth]
    {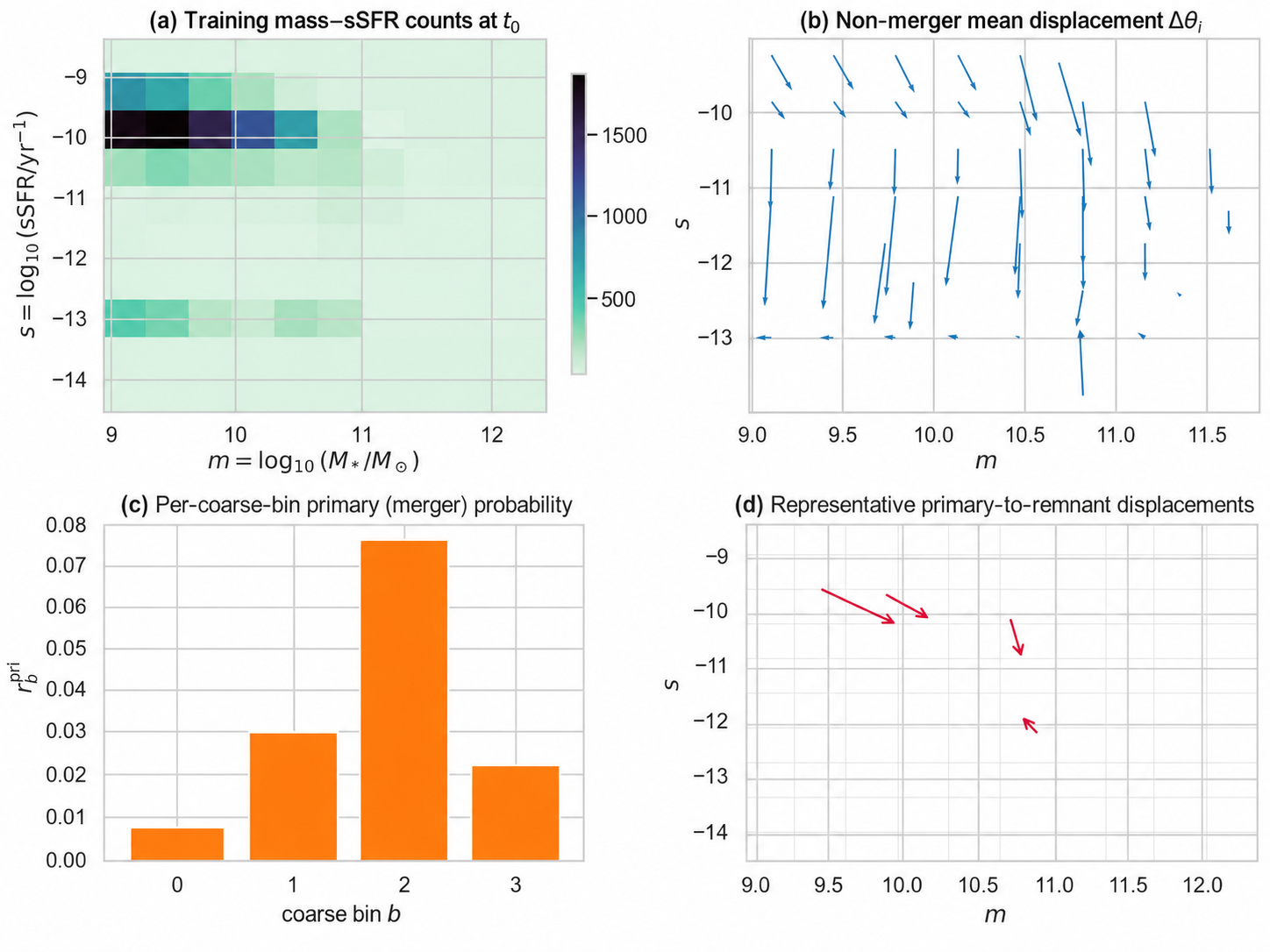}
    \caption{
    Finite-time effective dynamics estimated from the training root merger
    trees in the effective state defined by
    Equation~\eqref{eq:tng_effective_state}. The upper-left panel shows the
    initial training counts after mapping the $44$-cell adaptive partition back
    to the original $10\times10$ plotting grid, and the upper-right panel shows
    the conditional mean displacement of the non-merger channel on the same
    display grid. The lower panels show the primary-channel probability in the
    four coarse source bins and representative primary-to-remnant
    displacements. For the non-merger panel, the displacement components are
    plotted directly in $(m,s)$ data units with no additional multiplicative
    rescaling. In the primary-to-remnant panel, each arrow begins at the mean
    $t_0$ position of the primary progenitors in a coarse bin and ends at the
    weighted mean of the remnant-cell representative points under
    $\widehat{\overline P}_{\cdot b}^{\rm pri}$. Every realized
    adaptive cell is displayed, with no additional occupancy threshold beyond
    that used to construct the partition. The displayed primary-channel
    probabilities are plug-in estimates from the training split; uncertainty
    intervals are not shown. The arrows summarize evolution over the full
    snapshot interval and are neither instantaneous velocities nor isolated
    causal merger responses.
    }
    \label{fig:tng_effective_dynamics}
\end{figure}

Figure~\ref{fig:tng_effective_dynamics} shows modest mean stellar-mass growth and declining sSFR over much of the star-forming sequence in the non-merger channel. 
The horizontal concentration at $s=s_{\rm zero}$ represents systems with zero instantaneous SFR; positive but extremely small SFR values may lie below this coding value. 
The primary-to-remnant summaries generally shift toward larger stellar mass and, for initially star-forming coarse bins, lower sSFR, while marginalizing over partner state, merger time, event-level variables, and post-merger evolution. 
The largest displayed primary-channel probability is approximately $0.076$; this illustrates the state conditioning retained by the reduced jump channel for this simulation interval.

\subsection{Held-out prediction of the later population}
\label{subsec:tng_heldout_prediction}

Let $\bm n_0^{\rm test}$ and $\bm n_1^{\rm test}$ be the initial and realized
later count vectors of the held-out root merger trees. We compare
\begin{align}
    \widehat{\bm n}_1^{\rm static}
    &=
    \bm n_0^{\rm test},
    \notag\\
    \widehat{\bm n}_1^{\rm NM}
    &=
    \widehat{\mathsf Q}^{(0)}
    \bm n_0^{\rm test},
    \notag\\
    \widehat{\bm n}_1^{\rm T+J}
    &=
    \widehat{\mathsf A}
    \bm n_0^{\rm test}.
    \label{eq:tng_prediction_models}
\end{align}
The first model leaves the initial population unchanged. The second is a finite-time, count-preserving non-merger-only counterfactual constructed from $\widehat{\mathsf Q}^{(0)}$: it applies the conditional one-galaxy kernel to the entire initial cohort while suppressing resolved merger channels and is distinct from the infinitesimal transport field $\bm v$.
The third model retains both primary-to-remnant redistribution and secondary-progenitor loss. 
Only the transport-plus-jump model can reproduce the resolved merger contribution to the total-number balance.

\begin{figure}[t]
    \centering
    \includegraphics[width=\linewidth]
    {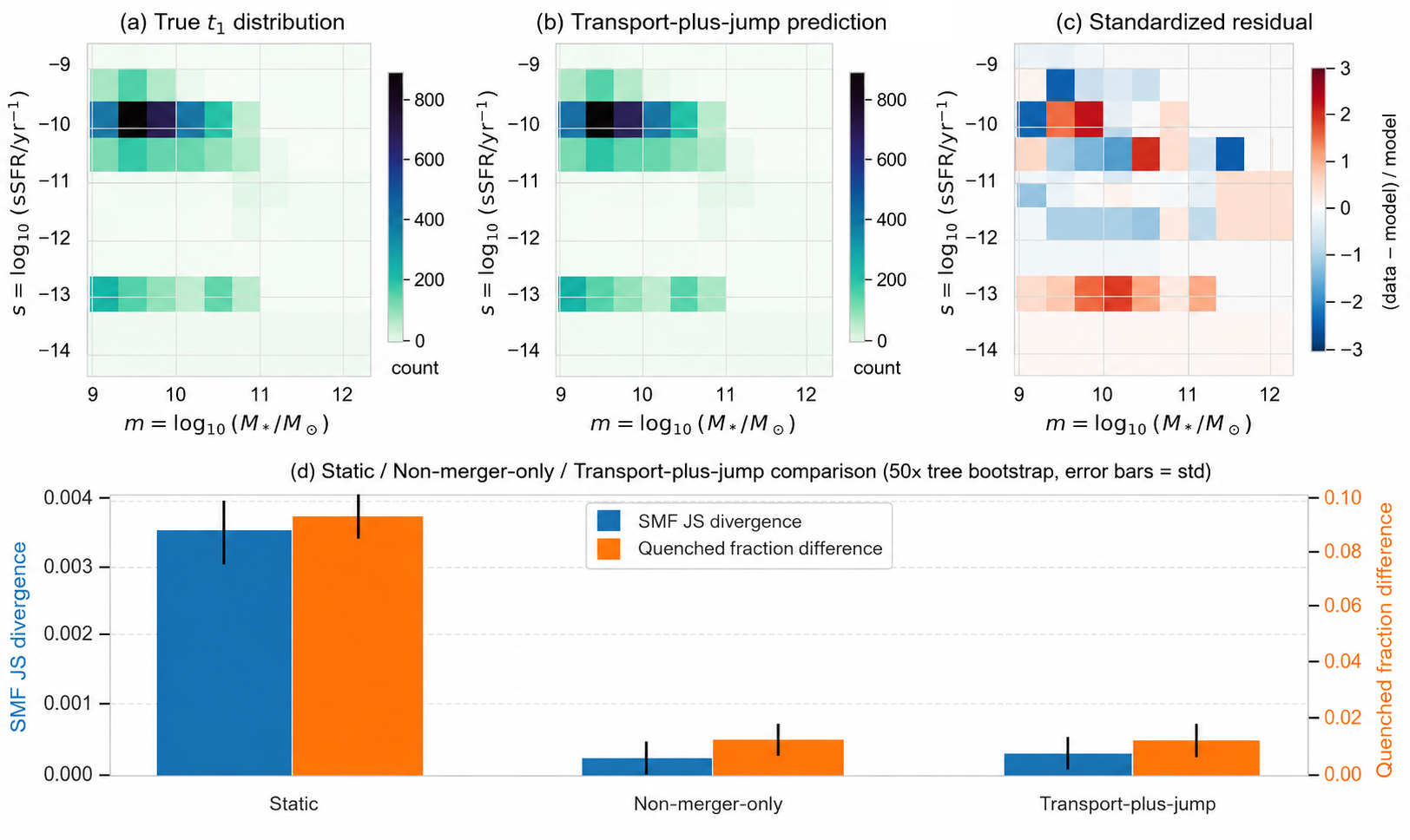}
    \caption{
    Held-out forward prediction from the initial population of the held-out
    root merger trees. The upper panels show the realized later distribution,
    the transport-plus-jump prediction in
    Equation~\eqref{eq:tng_prediction_models}, and the regularized Pearson-type
    residual
    $R_i=(n_{1,i}^{\rm test}-\widehat n_{1,i}^{\rm T+J})/
    \sqrt{\max(\widehat n_{1,i}^{\rm T+J},1)}$. The lower panel compares the
    static, non-merger-only, and transport-plus-jump models over 50 held-out
    root-tree bootstrap resamples conditional on the operator calibrated once
    from the training split. The blue bars show the stellar-mass-function
    Jensen--Shannon divergence, while the orange bars show the absolute
    quenched-fraction difference; the two-dimensional mass--sSFR divergence
    quoted in the text is a separate diagnostic. Error bars show the standard
    deviation over the held-out bootstrap resamples. Throughout these
    resamples, $\widehat{\mathsf Q}^{(0)}$, $\widehat{\mathsf A}$, and the
    adaptive and coarse partitions remain fixed. Whole root trees are
    resampled, so the primary, secondary, and remnant associated with a
    resolved event always move together.
    }
    \label{fig:tng_heldout_prediction}
\end{figure}

For non-negative count vectors $\bm p$ and $\bm q$, we normalize each vector
to unit sum and compute the base-2 Jensen--Shannon divergence
\begin{align}
    D_{\rm JS}(\bm p,\bm q)
    &=
    \frac{1}{2}
    D_{\rm KL}^{(2)}(\bm p\Vert\bm m)
    +
    \frac{1}{2}
    D_{\rm KL}^{(2)}(\bm q\Vert\bm m),
    \qquad
    \bm m
    =
    \frac{\bm p+\bm q}{2},
    \notag
\end{align}
where $D_{\rm KL}^{(2)}$ uses base-2 logarithms. Zero components are handled
by the standard limiting convention, with a numerical floor of $10^{-12}$
used to stabilize normalization and denominators. The two-dimensional forward and recovery
comparisons use the $44$ adaptive-cell count vectors. The stellar-mass-function
comparison uses a fixed 14-bin mass aggregation defined from the training
$t_0$ masses.

Figure~\ref{fig:tng_heldout_prediction} shows that the two finite-time
evolution models reproduce the star-forming ridge and the zero-SFR-coded
component much more accurately than the static baseline. For the normalized
two-dimensional mass--sSFR distribution, the Jensen--Shannon divergence
decreases from $0.034$ for the static baseline to $0.0033$ and $0.0035$ for
the non-merger-only and transport-plus-jump models, respectively. The absolute
quenched-fraction error decreases from $0.083$ to $0.011$ and $0.010$. The
separate stellar-mass-function diagnostic in the lower panel shows the same
qualitative improvement. The small differences between the two finite-time
evolution models do not establish a resolved shape-level ranking.

The merger-specific result is instead the number balance. The realized later
held-out count is $N_1^{\rm test}=5574$. Because both the static and
non-merger-only models preserve the initial count, they predict
$\widehat N_1^{\rm static}=\widehat N_1^{\rm NM}=5724$ and overpredict the
realized later count by approximately $2.7\%$. The sub-stochastic
transport-plus-jump operator predicts $\widehat N_1^{\rm T+J}=5564.7$ and
reduces the absolute count error to approximately $0.17\%$ by accounting for
disappearing secondaries.

\subsection{State recovery from noisy and censored observations}
\label{subsec:tng_state_recovery}

To connect the forward test to the inverse problem, we take the held-out later
states as latent truth and generate mock measurements according to
\begin{align}
    m_{{\rm obs},a}
    &=
    m_a+\epsilon_{m,a},
    &
    \epsilon_{m,a}
    &\sim
    \mathcal N(0,0.15^2),
    \notag\\
    s_{{\rm obs},a}^{\ast}
    &=
    s_a+\epsilon_{s,a},
    &
    \epsilon_{s,a}
    &\sim
    \mathcal N(0,0.30^2),
    \notag\\
    \delta_a
    &=
    \mathbbm{1}
    \left\{
        s_{{\rm obs},a}^{\ast}\geq s_{\rm lim}
    \right\},
    &
    s_{\rm lim}
    &=
    -11.2.
    \label{eq:tng_mock_observation}
\end{align}
The mock observational degradation is generated once with fixed random seed
44. The state-recovery experiment contains
$N_{\rm test}=N_1^{\rm test}=5574$ held-out later galaxies. The held-out
cohort is taken as given and the analysis is conditioned on its observed size.
This subsection is therefore an object-level, shape-conditional special case
of the catalog model in Section~\ref{subsec:catalog_likelihood}, with complete
sample selection and no additional Poisson count factor. The number-changing
prediction has already been tested in the preceding subsection.

Let $d_a$ denote the record for object $a$. When $\delta_a=1$, the measured
value $s_{{\rm obs},a}^{\ast}$ is retained; when $\delta_a=0$, the record
contains only the upper-limit event
$s_{{\rm obs},a}^{\ast}<s_{\rm lim}$. For a candidate state $(m,s)$, the
object-level likelihood is
\begin{align}
    \ell_a(d_a\mid m,s)
    &=
    \mathcal N
    \left(
        m_{{\rm obs},a}\mid m,0.15^2
    \right)
    \begin{cases}
        \mathcal N
        \left(
            s_{{\rm obs},a}^{\ast}\mid s,0.30^2
        \right),
        & \delta_a=1,\\
        \Phi
        \left(
            \dfrac{s_{\rm lim}-s}{0.30}
        \right),
        & \delta_a=0,
    \end{cases}
    \notag
\end{align}
where $\Phi$ is the standard normal cumulative distribution function. Thus,
non-detections remain censored records rather than point measurements at the
limit.

Let $\theta_i=(m_i,s_i)$ be the representative point of adaptive cell $C_i$.
The three priors are
\begin{align}
    \pi_i^{\rm flat}
    &=
    \frac{1}{N_{\rm cell}},
    \notag\\
    \pi_i^{\rm static}
    &=
    \frac{
        n_{0,i}^{\rm train}
    }{
        \displaystyle\sum_j n_{0,j}^{\rm train}
    },
    \notag\\
    \pi_i^{\rm dyn}
    &=
    \frac{
        \left(
            \widehat{\mathsf A}
            \bm n_0^{\rm train}
        \right)_i
    }{
        \displaystyle
        \sum_j
        \left(
            \widehat{\mathsf A}
            \bm n_0^{\rm train}
        \right)_j
    }.
    \label{eq:tng_mock_prior_definitions}
\end{align}
The flat prior is the weak baseline and assigns equal probability to each of
the 44 adaptive cells. It is therefore grid dependent and should not be
interpreted as a uniform density with respect to area in the $(m,s)$ plane.
The prior labeled ``static'' is specifically the frozen, unpropagated training
distribution at the initial epoch; it does not use the true held-out population
at the later epoch. The dynamical prior propagates the same training population
through the independently calibrated reduced operator. This construction
isolates the contribution of cross-epoch propagation while keeping the state
grid and object-level likelihood fixed.

For prior $k\in\{{\rm flat},{\rm static},{\rm dyn}\}$, the discrete posterior
and its population stack are
\begin{align}
    p_{ai}^{(k)}
    &=
    \frac{
        \ell_a(d_a\mid\theta_i)
        \pi_i^{(k)}
    }{
        \displaystyle
        \sum_j
        \ell_a(d_a\mid\theta_j)
        \pi_j^{(k)}
    },
    \notag\\
    \widehat n_{1,i}^{{\rm stack},(k)}
    &=
    \sum_{a\in{\rm test}}
    p_{ai}^{(k)}.
    \label{eq:tng_posterior_stack}
\end{align}
The stack retains each galaxy's state uncertainty instead of replacing its
posterior by a point estimate. Because each object posterior is normalized and
the analysis conditions on the held-out cohort size,
$\sum_i\widehat n_{1,i}^{{\rm stack},(k)}=N_{\rm test}$ for every prior.

\begin{figure}[t]
    \centering
    \includegraphics[width=\linewidth]{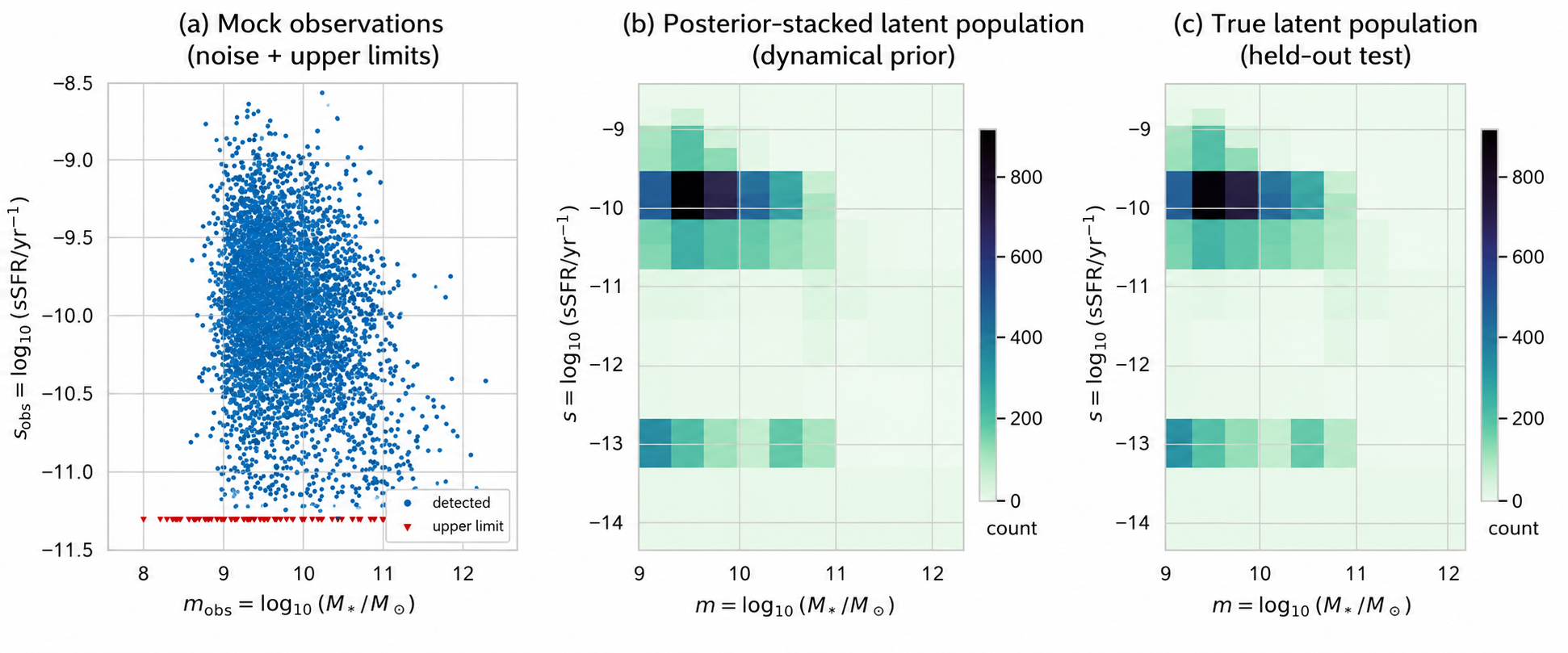}
    \caption{
    Recovery of the held-out latent later population after observational
    degradation. The left panel shows noisy measurements and sSFR upper limits
    from Equation~\eqref{eq:tng_mock_observation}; upper limits are displayed
    at the common plotting threshold and are not treated as measured values at
    that threshold. The middle panel is the posterior stack under the
    dynamical prior, and the right panel is the true latent population
    distribution.
    }
    \label{fig:tng_state_recovery}
\end{figure}

Figure~\ref{fig:tng_state_recovery} shows that both population-informed priors
improve substantially over the flat prior. Relative to the frozen-static
prior, dynamical propagation reduces the population Jensen--Shannon divergence
from $0.0048$ to $0.0021$. 
The reported individual-state root-mean-square errors (RMSEs) remain nearly unchanged: the stellar-mass RMSE remains $0.15$ dex, while the sSFR RMSE changes only from $0.37$ to $0.36$ dex, indicating that the additional gain from dynamical propagation is concentrated in the later-epoch latent-population reconstruction.
These diagnostics are computed from the single mock realization generated with seed 44.

\subsection{Scope and outcome}
\label{subsec:tng_scope_limitations}

The demonstration uses one simulation realization, one snapshot interval, a
fixed two-dimensional effective physical state, and empirical grid operators.
The minimum workflow fixes $(m,s)$ in advance and does not compare it with an
enlarged state containing halo, environmental, gas, or assembly variables.
Consequently, the experiment does not establish predictive adequacy of the two-dimensional state, and one interval cannot test the Chapman--Kolmogorov relation. 
Interpretation remains conditional on the adopted discretization, sparse-event pooling, sample definition, and merger-tree conventions.

The held-out root-tree bootstrap quantifies evaluation-sample variation conditional on the single operator fitted to the training split. 
Operator-estimation, model-specification, simulation, and survey-transfer uncertainties remain outside these conditional bootstrap intervals.

Within these restrictions, the calculation establishes three points. 
A calibrated finite-time operator predicts the held-out later population much more accurately than a static baseline. 
The explicit reduced merger channel recovers the two-to-one number balance that a count-preserving one-galaxy kernel cannot represent. 
Finally, the propagated dynamical prior improves reconstruction of the latent later population from noisy and censored mock measurements relative to the frozen-static prior, while the reported individual-state RMSEs remain nearly unchanged. 
The proof of concept therefore supplies a concrete implementation of the observation-directed, simulation-assisted state-and-dynamics inference strategy.

\section{Conclusion and Outlook}
\label{sec:conclusion}

\subsection{Conclusion}
We have reformulated galaxy population evolution as an
observation-directed inference problem.  
The scientific target is the joint reconstruction of individual effective states, the population intensity over those states, and the operators that connect the population across cosmic time. 
The galaxy manifold is therefore an effective state space for a specified prediction task and temporal resolution.  
Its coordinates must be tested through predictive
adequacy.  
Measured variables $\bm{x}$, learned representations $\bm{z}$, and effective physical states $\theta$ remain distinct, even when they have the same dimension.

The intrinsic population is represented by a finite non-negative
measure, so its total intensity can evolve as well as its shape.  Its
dynamics combine one-galaxy transport, entry, and removal with a
nonlocal binary-merger channel.  Separating the merger-rate kernel from
the conditional remnant-state kernel distinguishes how often a pair
merges from where its remnant appears in the multivariate state space.
The corresponding finite-time construction is directly estimable from
snapshots and merger trees.  Its missing probability mass records
secondary progenitors that no longer survive as independent systems,
rather than a failure of normalization.

An observational kernel maps this intrinsic population to noisy, selected, and possibly censored catalogs.  Together with the evolutionary model, it defines a catalog likelihood and makes the transport--jump solution a time-dependent population prior for individual-galaxy state inference.  
Catalogs at different epochs can therefore be constrained jointly through one dynamically consistent model.
Projected summaries such as luminosity and stellar-mass functions remain useful diagnostics, but their induced
dynamics are generally not closed.  
The Schechter and downsizing examples show that agreement in a low-dimensional projection constrains
combinations of physical channels without, by itself, identifying those channels uniquely.

The inverse problem becomes estimable only after its exact structure and its uncertain assumptions are separated.  Positivity, normalization, progenitor symmetry, conservation conditions, and count balance should be imposed by construction.  
Smoothness, locality, and similarity to a reference simulation should instead enter as regularizing priors whose influence can be examined.  
Simulations provide controlled trajectories for testing state adequacy, calibrating reduced operators, and validating posterior recovery.
When the forward likelihood is intractable, simulation-based inference is a computational route to the posterior of this specified model.

The IllustrisTNG proof of concept implemented this chain using the fixed two-dimensional effective physical state of Equation~\eqref{eq:tng_effective_state}. 
Training root merger trees were used to calibrate the non-merger and reduced merger operators, which were then evaluated on independent held-out root merger trees.

Both the non-merger-only and transport-plus-jump predictions reproduced the normalized later mass--sSFR distribution substantially better than an unchanged static baseline. 
Their shape-level performance was comparable for the diagnostics considered, whereas the explicit merger
channel accounted for the loss of secondary progenitors and supplied the corresponding two-to-one number balance that a count-preserving one-galaxy transition cannot represent.

For the inverse test, the held-out later states were degraded with Gaussian measurement noise and censored sSFR information. 
Conditioning on the held-out cohort and its observed size, the dynamically propagated prior improved the posterior-stacked reconstruction of the latent population relative to the frozen-static prior, while the reported individual-state RMSEs remained nearly unchanged.
These results provide a concrete implementation of finite-time operator calibration, held-out population prediction, and posterior reconstruction of the latent population under a known observation model.

\subsection{Outlook}
An observational application should begin by defining the prediction
task and testing whether the proposed state retains the transition
information required for it.  The survey-specific observational kernel
must then propagate detection, deblending, measurement uncertainty,
redshift uncertainty, and censoring without confusing them with
intrinsic population evolution.  
Multiple epochs and correlated observables should be analyzed jointly, with operator uncertainty, cosmic variance, and simulation-to-survey discrepancy carried into both population forecasts and individual-state posteriors.  
Measurements that respond selectively to mergers, quenching, environment, gas, or structure will be especially valuable for reducing channel degeneracies.

The framework provides a reduced and testable interface through which the population-level consequences of galaxy-formation processes can be compared and constrained by observations.
The galaxy manifold thereby serves as an effective state space for uncertainty-quantified inference of galaxy population evolution.

\acknowledgments

We would like to express our deep gratitude to Shiro Ikeda, Kenji Fukumizu, and Satoshi Kuriki for valuable discussions and insightful comments on this research.
This work was supported by JSPS Grant-in-Aid in Scientific Research (24H00247), JST CREST Grant No. JPMJCR24Q1, and by the Joint Research program of the Institute of Statistical Mathematics (General Research 2) entitled ``Machine-Learning Cosmogony: From Structure Formation to Galaxy Evolution.''



\bibliographystyle{JHEP}
\bibliography{manifold_learning}

\end{document}